\documentclass[%
 reprint,
superscriptaddress,
 amsmath,amssymb,
 aps,
]{revtex4-2}

\usepackage{graphicx}
\usepackage{dcolumn}
\usepackage{bm}
\usepackage{amsmath}
\usepackage{mathtools}
\usepackage{physics}
\usepackage{ulem}
\usepackage{xcolor}
\usepackage{braket}
\usepackage{subcaption} 
\usepackage{ragged2e} 
\DeclareCaptionJustification{justified}{\justifying}
\usepackage{comment}

\newcommand{\RE}{Rényi }

\newcommand{\iu}{{\rm i}}

\usepackage{hyperref}
\hypersetup{
	colorlinks=true,
	linkcolor=blue,
	citecolor=blue,
	filecolor=magenta,
	urlcolor=cyan
}

\begin{document}

\preprint{APS/123-QED}

\title{Purification Dynamics in a Continuous-time Hybrid Quantum Circuit Model}

\def\TCM{{T.C.M. Group, Cavendish Laboratory, JJ Thomson Avenue, Cambridge CB3 0HE, United Kingdom}}
\def\oxf{{Rudolf Peierls Centre for Theoretical Physics, Clarendon Laboratory, Parks Road, Oxford OX1 3PU, United Kingdom}}
\def\UCL{{London Centre for Nanotechnology, University College London,
Gordon St., London WC1H 0AH, United Kingdom}}

\author{Sebastian Leontica}
\affiliation{\UCL}
\affiliation{\oxf}
\author{Max McGinley}
\affiliation{\TCM}
\affiliation{\oxf}

\date{\today}

\begin{abstract}

We introduce a continuous time model of many-body quantum dynamics based on infinitesimal random unitary operations, combined with projective measurements. We consider purification dynamics in this model, where the system is initialized in a mixed state, which then purifies over time as a result of the measurements. By mapping our model to a family of effective 1D quantum Hamiltonians, we are able to derive analytic expressions that capture how the entropy of the system decays in time. Our results confirm the existence of two distinct dynamical phases, where purification occurs over a timescale that is exponential vs.~constant in system size. We compare our analytic expressions for this microscopic model to results derived from field theories that are expected to capture such measurement-induced phase transitions, and find quantitative agreement between the two.

\end{abstract}

\maketitle

\section{Introduction}

The continuing development of programmable quantum devices with increasing numbers of degrees of freedom has led to a great deal of interest in addressing fundamental questions regarding the dynamics of information in many-body quantum systems \cite{Calabrese2005, Kim2013, Ho2017, Calabrese2009, Hosur2016, Nahum2017,Zhou2019, Chan2018, Roberts2017, Nahum2018, Keyserlingk2018, Rakovszky2018}. In recent years, there has been a particular focus on the competition between unitary operations, which generate entanglement, and local projective measurements, which are non-unitary processes that break entanglement. Models of dynamics that feature both of these ingredients are often referred to as hybrid quantum circuits, the study of which has led to the discovery of a sharp entanglement phase transition driven by the rate of measurements, separating regimes where many-body entanglement is either stable or fragile against these measurements \cite{Li2018,Skinner2019,Chan2019, Li2019, Bao2020, Choi2020, Jian2020, Zabalo2020, Nahum2021, Ippoliti2021, Zabalo2022, Friedman2022}. Typically, the studied geometry is that of a 1D chain of qudits, but similar transitions have also been found in more complex geometries such as random tensor networks \cite{Vasseur2019, Hayden2016, Bao2019, Yang2022}.

The existence of this transition was first understood in terms of the entanglement structure of an ensemble of pure many-body states at equilibrium. Subsequent studies also revealed the existence of a simultaneous dynamical phase transition, which can be understood as the ability of the measurement protocol to learn an initially mixed state \cite{Gullans2020, Gullans2020(1), Gopalakrishnan2021, Noel2022}. The latter suggests a connection between the dynamics of hybrid quantum circuits and quantum error correcting codes \cite{Fan2021}, which by construction protect information against deleterious non-unitary processes. The transition was also shown to play an important role in the context of simulating the behaviour of open quantum systems \cite{Vovk2022, Azad2023, Chen2023, Cheng2023}.

These considerations have led to the notion of purification dynamics, where one studies how the entropy of an initially mixed state decreases over time as a result of the measurements. Away from the critical measurement rate we find two phases where the state purifies over a timescale that increases exponentially with system size (`mixed phase') or is independent of the system size (`purifying phase'). To understand the phenomenology of these phases in a fully quantitative way, arguments based on capillary wave theory have been put forward \cite{Li2021}. Using an effective field theory which is expected to capture the universal features of the transition, one can obtain concrete predictions of how the purity of the system will depend on time in each phase. However, direct verification of these predictions by means of a direct calculation from a microscopic model are as of yet lacking.
 
In this paper, we introduce and study a hybrid quantum circuit model of dynamics that is defined in continuous time, the properties of which we are able to calculate analytically. In particular, by means of a mapping onto an effective Hamiltonian, we are able to compute the time dependence of a particular family of operator-space entanglement measurements, which can be related to the purity of the system at a time $t$, starting from a maximally mixed initial state. We look in detail at both the mixed and purifying phases, as well as at the transition between them. Our results agree with those of capillary wave theory in both phases: the entropy decays exponentially with time in the purifying phase, and decreases as $-\log t$ in the mixed phase over an exponentially long time window [Eqs.~(\ref{eq:SLogDecreasePBC},\ref{eq:entropyearly})]. In our calculations, we consider both periodic and open boundary conditions, and show that the two choices give rise to quantitatively different behaviour when in the mixed phase: in particular, a $(1/2)\log N$ contribution to the entropy appears when we impose periodic boundary conditions, but this is absent for open boundary conditions. We also look at the dynamics at criticality, where there exists a regime during which the entropy decays algebraically, Eq.~\eqref{eq:EntropyCriticalAlgebraic}.

The structure of our paper is as follows. In Section \ref{sec:unitmodel} we introduce a continuous time model of dynamics based on infinitesimal random unitary operations, and describe how one can calculate various measures of entanglement and information spreading in this model. We supplement the unitary dynamics with projective measurements in Section \ref{sec:measurements}, and in Section \ref{sec:fermionic}, we explain how the resultant unitary-projective dynamics can be mapped onto imaginary-time evolution under an effective 1D Hamiltonian. We then present our main quantitative results in Section \ref{sec:QuantitativeResults}, giving analytic expressions that quantify how the purity of the system increases as a function of time in the purifying/mixed phase and at criticality. Finally, we discuss our results and conclude in Section \ref{sec:Discussion}.



\section{Continuous-time random circuit model}
\label{sec:unitmodel}


In this section, we introduce a random unitary circuit (RUC) model of unitary dynamics, and describe how its entanglement properties can be analysed. We will later incorporate measurements into this model, which will allow us to study the dynamics of purification.

\begin{figure}[t!]
    \includegraphics[width=0.47\textwidth]{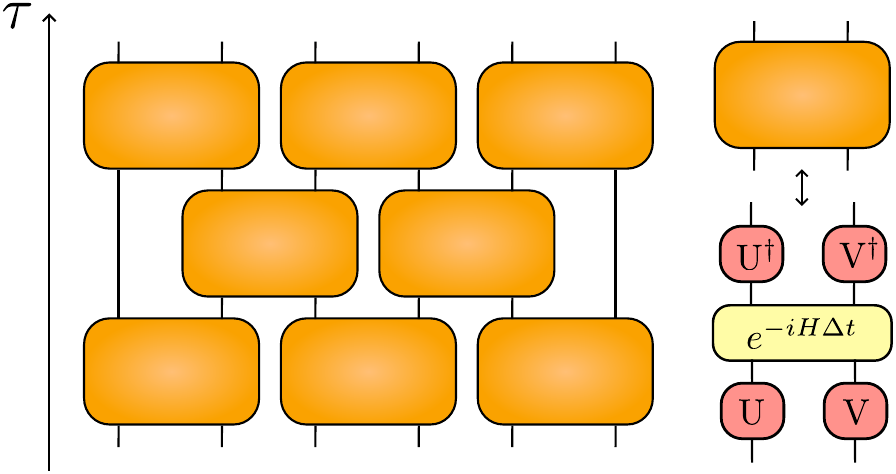}
    \caption{Schematic representation of the random circuit geometry for open boundary conditions. The construction of the unit cells is illustrated on the right. The Hamiltonian $H$ and evolution time $\Delta t$ are kept fixed, but the random unitaries $U, V$ are sampled independently at each spacetime location in the circuit.}
    \label{fig:brickwork}
\end{figure}

We consider a one-dimensional array of $N$ qudits, each with a local Hilbert space of dimension $d$. The evolution is driven by a spatially local unitary circuit with a brickwork structure, illustrated in Fig.~\ref{fig:brickwork}. In a given timestep  $\tau = 1, 2, \ldots$, two-site unitaries are applied to each pair of qudits on the odd bonds $(2j-1, 2j)$, followed by another layer of unitaries on the even bonds $(2j, 2j+1)$. These elementary two-site unitaries each have the same structure, also depicted in Fig.~\ref{fig:brickwork}. First, single-site gates $U\otimes V$ are applied, followed by evolution under some two-qudit Hamiltonian $H$ for a time $\Delta t$, and finally the change of basis is undone by applying the inverse single-site rotations $U^{\dagger}\otimes V^{\dagger}$. We denote the unitary operator describing the evolution from time 0 to $\tau$ as $W(\tau)$.

Throughout this work, $H$ will be treated as a free parameter of the model and it is kept fixed across both time and space. To simplify calculations, we will assume it is real, hermitian and symmetric under swapping the two qudits it acts on. The single-qudit unitaries will be sampled randomly and independently from the Haar ensemble for each unit cell. We will generally be interested in the limit where $\Delta t \rightarrow 0$, which we refer to as the continuous-time limit. Note that the state only evolves by an infinitesimal amount in each timestep, in contrast to discrete-time RUC models of quantum dynamics (e.g.~Refs.~\cite{Nahum2017, vonKeyserlingk2018}). A model of continuous-time dynamics was studied numerically in \cite{Szyniszewski2020}. Our method for constructing the unit cell is more general and more easily amenable to analytical treatment.

Our focus will be on the dynamics of entanglement and quantum information in these continuous-time models. For this purpose, it is useful to consider the Choi-Jamiolkowski (CJ) state $\ket{W(\tau)}$ corresponding to the unitary $W(\tau)$. This state is defined on two copies of the system, which we can associate with the inputs and outputs of the time evolution operator. Formally, we have  $\ket{W(\tau)} = [\mathbb{I} \otimes W(\tau)]\ket{\Phi^+}$, where $\ket{\Phi^+} = \bigotimes_{j=1}^N (d^{-1/2}\sum_{a=1}^d \ket{a} \otimes \ket{a})$ consists of maximally entangled states between each input qudit and its corresponding output \cite{Jiang2013}. Many important quantities that are used to diagnose the spreading of quantum information can be expressed as simple functions of this operator-state $\ket{W(\tau)}$ \cite{Hosur2016}, and in particular we will find this representation useful when it comes to studying purification dynamics.

As is now common in studies of RUC dynamics,  we use the \RE entropies to quantify the entanglement properties of the state $\ket{W(\tau)}$
\begin{equation}
S^{(n)}(\rho_A)=\frac{1}{1-n} \log \tr\left(\rho_A^{n}\right),
\label{eq:renyi}
\end{equation}
where $n$ is some positive parameter. Here, $\rho_A$ is the reduced density matrix of $\ket{W(\tau)}$ corresponding to some subset $A$ of inputs and outputs. Compared to the usual von Neumann entropy $S_{\rm vN}$, the \RE entropies for $n = 2, 3, \ldots$ are more amenable to analytic studies, since they only involve integer moments of the density matrix and hence can be computed using a replica method. The von Neumann entanglement entropy $S_{\rm vN}$ can be recovered by constructing an analytical continuation of the function and taking the limit $n \to 1$ (see, e.g.~Ref.~\cite{Bao2022}).

For the largest part of this work, we will only be concerned with the second \RE entropy $S^{(2)}(\rho_A)$, which is the simplest to evaluate. This is a lower bound on the von Neumann entropy $S_{\rm vN} \geq S^{(2)}$, which in certain cases is known to be asymptotically tight \cite{Bianchi2018}. Since the purity $ \Tr[(\rho^A)^2] \equiv \exp[-S^{(2)}(\rho^A)]$ is a quadratic function of $\ket{W(\tau)}\bra{W(\tau)}$, it can be expressed using a fourfold copy of the evolution operator, which we denote
\begin{align}
    {\mathbf W}^{(2)}(\tau) \coloneqq (W(\tau) \otimes W^*(\tau))^{\otimes 2}.
    \label{eq:DuplicatedUnitary}
\end{align}

Note here that the operator replicated in the expression differs from $\ket{W(\tau)}\bra{W(\tau)}$ by a reshuffling of the indices. Henceforth, we will use this convention, but retain the essence of the CJ isomorphism by noting that we treat inputs and outputs on par when discussing \RE entropies.

Define (unnormalized) states
\begin{align}
    \ket{\mathbf{I}}_j &= \sum_{a,b=1}^d \Big(\ket{a}\otimes\ket{a}\otimes \ket{b}\otimes \ket{b}\Big)_j,\label{eq:StateI} \\
    \ket{\mathbf{S}}_j &= \sum_{a,b=1}^d \Big(\ket{a}\otimes\ket{b}\otimes \ket{b}\otimes \ket{a}\Big)_j, \label{eq:StateS}
\end{align}
which live in the fourfold-replicated Hilbert space of each physical site $j$. In terms of these, we have
\begin{align}
    e^{-S^{(2)}} = \Tr[(\rho^A)^2] = \braket{\Psi_{A_\mathrm{out}}| {\mathbf W}^{(2)}(\tau) | \Psi_{A_\mathrm{in}}},
    \label{eq:entropy}
\end{align}
where we denote the set of input (output) sites included in the region $A$ as $A_{\rm in}$ ($A_{\rm out}$), and the states
\begin{align}
    \ket{\Psi_{A_\mathrm{in}}} = \left(\bigotimes_{j \in A_\mathrm{in}}\ket{\mathbf{S}}_j \right)\otimes\left(\bigotimes_{j \notin A_\mathrm{in}}\ket{\mathbf{I}}_j\right),
\end{align}
and similar for $\ket{\Psi_{A_\mathrm{out}}}$.

To make progress, we look at the average of the \RE entropy \eqref{eq:entropy} over the random ensemble of unitary circuits. More precisely, we will evaluate the average purity as opposed to the average entropy, which is equivalent to performing averages inside the logarithm of Eq.~\eqref{eq:renyi}. This simplification is common in analyses of RUCs \cite{Bao2020}, and still recovers the correctly-averaged von Neumann entropy if one takes the replica limit $n\to 1$. Accordingly, averaging the purity amounts to replacing ${\mathbf W}^{(2)}(\tau)$ with its ensemble average $\overline{{\mathbf W}^{(2)}(\tau)}$. As shown in App. \ref{app:diagrammatic}, $\overline{{\mathbf W}^{(2)}(\tau)}$ maps states spanned by tensor products of $\ket{\mathbf{I}}_j$, $\ket{\mathbf{S}}_j$ to other such states, meaning we can focus on the restriction of this averaged operator to the subspace $V(S_2)^{\otimes N}$, where $V(S_2) = \text{span}(\ket{\mathbf{I}}, \ket{\mathbf{S}})$. 

Because the single-site Haar-random unitaries appearing in each of the two-site elementary blocks  of the circuit [Fig.~\ref{fig:brickwork}(b)] are sampled independently, we can consider the ensemble average of the evolution under a single one of these blocks, which we denote $\mathcal{T}:V(S_2)^{\otimes 2}\to V(S_2)^{\otimes 2}$. Using the Weingarten diagrammatic calculus as seen in App. \ref{app:diagrammatic}, we find that, for small $\Delta t$, this map can be expressed as
\begin{equation}
\label{eq:transfunitcell}
    \mathcal{T}_{ij} = 1 - \Delta t^2 \Omega(H)\mathcal{W}g^{(i)}\mathcal{W}g^{(j)} \left(1-\sigma_z^{(i)}\sigma_z^{(j)}\right)+\mathcal{O}(\Delta t ^4),
\end{equation}
where $i,j$ label the sites on which the unit cell acts, $\Omega(H)$ is a measure of the entangling power of the Hamiltonian
\begin{equation}
\label{eq:omega}
    \Omega(H) = d^2 \operatorname{tr}(H^2)-2d\operatorname{tr}(\operatorname{tr}_1(H)^2)+\operatorname{tr}(H)^2,
\end{equation}
and $\mathcal{W}g$ is the Weingarten matrix corresponding to the symmetric group $S_2$
\begin{equation}
    \mathcal{W}g = \frac{1}{d(d^2-1)}
    \begin{bmatrix}
    d & -1 \\
    -1 & d
    \end{bmatrix} = \frac{1}{d^2-1}\left(1-\frac{\sigma_x}{d}\right).
\end{equation}

The induced evolution can be equivalently described using the effective imaginary-time Hamiltonian
\begin{equation}
\label{eq:hamiltonianij}
    \mathcal{H}_{ij} = \Omega(H)\mathcal{W}g^{(i)}\mathcal{W}g^{(j)} \left(1-\sigma_z^{(i)}\sigma_z^{(j)}\right),
\end{equation}
in terms of which the unit cell map is
\begin{equation}
    \mathcal{T}_{ij} = e^{-\Delta t^2\mathcal{H}_{ij}} + \mathcal{O}(\Delta t ^4).
    \label{eq:Tij}
\end{equation}

It is interesting to note that there are no contributions from odd powers of $\Delta t$ in the expansion of Eq. \ref{eq:transfunitcell}. Looking at the form of the Hamiltonian \eqref{eq:hamiltonianij}, we see that in the effective Hilbert space spanned by the states (\ref{eq:StateI}, \ref{eq:StateS}), the only mobile degrees of freedom are domain walls separating regions of $\ket{\mathbf{I}}$ from $\ket{\mathbf{S}}$, consistent with discrete-time RUCs discussed previously \cite{Nahum2017, vonKeyserlingk2018}. The initial Hamiltonian $H$ only enters the expression through its entangling rate $\Omega(H)$. This sets the overall timescale of quantum information transfer through the system. In App. \ref{app:multiplereplicas}, we compute the transfer matrix for a higher number of replicas and show that this statement holds more generally. This result suggests that the qualitative behavior of entanglement dynamics derived from our model should be insensitive to most of the microscopic details, and hence applicable to a wide range of physical processes.

The propagator for the whole circuit $\mathcal{T}$ can be constructed by concatenating the two-site maps \eqref{eq:Tij} according to the brickwork circuit structure illustrated in Fig.~\ref{fig:brickwork}. We have
\begin{equation}
    \mathcal{T}(\tau) = \left(\prod_{\substack{i=2 \\ i \, \mathrm{even}}}^{N-2} e^{-\Delta t^2 \mathcal{H}_{i,i+1}} \prod_{\substack{i=1 \\ i \, \mathrm{odd}}}^{N-1} e^{-\Delta t^2 \mathcal{H}_{i,i+1}}\right)^{\tau} + \mathcal{O}(\tau\Delta t ^4).
\end{equation}
We now define the effective time as $t = \tau\Delta t^2 $ and take the limit $\tau \to \infty$, $\Delta t \to 0$ such that $t$ is kept constant. Using the Suzuki-Trotter formula, we find the limit of the previous equation
\begin{equation}
    \mathcal{T}(t) = \exp\left(-t\sum_{i=1}^{N-1}\mathcal{H}_{i,i+1}\right).
\end{equation}
We reiterate here that this operator acts as the restriction of $\overline{{\mathbf W}^{(2)}(\tau)}$ to its invariant subspace $V(S_2)^{\otimes N}$, and therefore may replace it in average entropy calculations (e.g. averaging Eq. \ref{eq:entropy}).

In its current form, the effective Hamiltonian $\sum_{i=1}^{N-1}\mathcal{H}_{i,i+1}$ is not Hermitian, but can be made so through a local similarity transformation, a technique commonly encountered in the study of non-equilibrium dynamics \cite{Hinrischen2000}. If we define a new evolution operator by $\tilde{\mathcal{T}} = (\mathcal{W}g^{-\frac{1}{2}})^{\otimes N}\mathcal{T}(\mathcal{W}g^{\frac{1}{2}})^{\otimes N}$, each 2-local term in the effective Hamiltonian transforms as $\tilde{\mathcal{H}}_{ij} = (\mathcal{W}g^{-\frac{1}{2}})^{\otimes 2}\mathcal{H}_{ij}(\mathcal{W}g^{\frac{1}{2}})^{\otimes 2}$, which gives us the Hermitian interaction
\begin{equation}
\begin{split}
    \tilde{\mathcal{H}}_{ij} &= \frac{\gamma}{2}\left[ 1-\frac{d^2-1}{d^2}\sigma_z^{(i)}\sigma_z^{(j)}\right. \\ &+\frac{1}{d^2}\sigma_x^{(i)}\sigma_x^{(j)}-\left.\frac{1}{d}(\sigma_x^{(i)}+\sigma_x^{(j)})\right],
\end{split}
\end{equation}
where the overall strength is given by $\gamma =2\Omega(H)/(d^2-1)^2$. This type of local interaction is found in the literature both as the quantum equivalent of the classical two-dimensional axial next-nearest neighbor Ising model (ANNNI) \cite{Barber1981, Selke1988} or more recently as the Jordan-Wigner transform of the balanced interacting Kitaev chain \cite{Mahyaeh2020, Maiellaro2022}.

In the limit of $d\to \infty$ we are left with a simple ferromagnetic nearest-neighbor Hamiltonian, with each domain wall incurring an energy penalty of $\gamma$. The Hamiltonian is symmetric under the global spin-flip operator $\mathcal{C} = \prod_i \sigma_x^{(i)}$, as can be seen through the commutation relation $[\mathcal{C},\mathcal{H}_{ij}]=0$.


\section{Including measurements}
\label{sec:measurements}

In this section, we will introduce the formalism that can be used to incorporate measurements into the random circuit evolution. For the purpose of this work, we will consider projective measurements in the computational basis of each qudit that occur stochastically. The same framework can accommodate for weak-measurement schemes as seen in Ref.~\cite{Bao2020}. Due to the continuous nature of our circuits, the effective model will be identical in the two cases.

A projective measurement is a non-unitary stochastic process, where the wavefunction of the system $\ket{\psi}$ collapses to a post-measurement state $\ket{m}$ with probability $p_m = |\braket{m|\psi}|^2$. Here, the set of wavefunctions $\{\ket{m}\}$ is the computational basis in which the measurement is performed and $m = 1,\,2, \ldots d$. For any fixed realisation of the random unitary circuit and positioning of the measurements, the final state of the system will depend on all the measurement outcomes $\mathbf{m} = (m_1,\,m_2, \ldots )$. Thus, we can write the ensemble of final states as $\{(p_{\mathbf{m}}, \ket{W_\mathbf{m})}\}$, where $p_{\mathbf{m}}$ is the joint probability of the measurement results, and $\ket{W_\mathbf{m})}$ is the (normalized) conditional state. As before, we will imagine the Choi-Jamiolkowski state, so $\ket{W_\mathbf{m})}$ is a state on a twofold copy of the system, and is constructed by preparing a maximally entangled state between the two copies in the computational basis, and evolving one of the copies under the evolution in question.

As is typical in the study of hybrid quantum circuits, our interest is on the statistics of the entanglement properties of individual conditional wavefunctions $\ket{W_\mathbf{m})}$; see, e.g.~Refs.~\cite{Skinner2019, Gullans2020}. The natural quantity to consider for this purpose is the von Neumann entropy, $S_{\rm vN}(\rho^A_\mathbf{m})$, where $\rho^A_{\mathbf{m}}$ is the reduced density matrix of $\ket{W_{\mathbf{m}}}$ over a subset of inputs and outputs $A$.  Specifically, we would want to compute the average of this quantity over all realizations of the random circuit, measurement locations and measurement results, which we denote $\overline{S_{\rm vN}(\rho^A_{\mathbf{m}})}$. However, this quantity is very difficult to compute directly in random circuit models. We follow Ref.~\cite{Bao2020} and introduce the series of related quantities
\begin{equation}
\label{eq:ententropy}
    \Tilde{S}^{(n)}_A = \frac{1}{1-n} \log\abs{\frac{\sum_{\{M\}}p_{M}d^{\abs{M}(n-1)}\overline{\sum_{\mathbf{m}} p_{\mathbf{m}}^{n} \operatorname{tr}[(\rho_{\mathbf{m}}^{A})^n]}}{\sum_{\{M\}}p_{M}d^{\abs{M}(n-1)}\overline{\sum_{\mathbf{m}} p_{\mathbf{m}}^{n}}}},
\end{equation}
where ${M}$ labels a particular configuration of measurement locations in spacetime, which occurs with probability $p_M$, and $\mathbf{m}$ runs over all measurement results for the given configuration $M$. These quantities are related to measurement-averaged \RE entropies, with the main difference that each outcome is weighted by $p_{\mathbf{i}_M}^{n}$. The additional factor of $d^{\abs{M}(n-1)}$ ensures that the correct order of magnitude, in powers of $d$, of the correct weight is preserved, and only deviations from it are amplified by the number of replicas. The renormalization is also performed on average, i.e. we compute the average of the numerator and the denominator independently. Knowledge of this quantity for all integers $n\geq 2$ can be used to recover the average entanglement entropy $\overline{\tilde{S}_A}$ of subsystem $A$ using the replica limit
\begin{equation}
\overline{S_{\rm vN}(\rho^A_\mathbf{i})}=\lim _{n \rightarrow 1} \tilde{S}^{(n)}_A.
\end{equation}

Each term in the sums over $M$ appearing in the numerator and the denominator in Eq.~\eqref{eq:ententropy} is a scalar that depends linearly on the tensor
\begin{equation}
    \overline{\langle \mathbf{W}^{(n)} \rangle} = d^{\abs{M}(n-1)}\overline{\sum_{\mathbf{m}}p_{\mathbf{m}}^n \boldsymbol \left(W_m\otimes W_m^*\right)^{\otimes n}},
    \label{eq:ReplicatedTensorMeasurements}
\end{equation}
which is analogous to the duplicated state in Eq.~\eqref{eq:DuplicatedUnitary} defined earlier. The angled brackets are a short-hand notation for the weighted sum on the RHS. For simplicity, we once again revert to the normal operator indices, but keep in mind that the probabilities are obtained from expectation values of the appropriate projectors in the CJ state $\ket{W_m}$.

To see how this tensor evolves as the circuit progresses, let us consider how $\overline{\langle \mathbf{W}^{(n)} \rangle}$ is updated when a new measurement is performed on site $i$, the outcome of which we denote $m$. Each of the $n$ replicas transforms via the action of a projector $P_m^{(i)}$, which corresponds to the $m$th computational basis state for qudit $i$. Since all measurement outcomes $m$ are summed over in Eq.~\eqref{eq:ReplicatedTensorMeasurements}, we find
\begin{equation}
    \overline{\langle \mathbf{W}^{(n)} \rangle} \to d^{n-1}\mathcal{M}_i(\overline{\langle \mathbf{W}^{(n)} \rangle}) = d^{n-1}\sum_m (P_m^{(i)})^{\otimes 2n}\overline{\langle \mathbf{W}^{(n)} \rangle}.
\end{equation}

The effect of post-selection is included by assuming a perfect correlation of the measurement results in all $n$ replicas.

Since adding an infinitesimal time evolution to the averaged tensor only leads to linear transformations by left multiplication due to both the chaotic dynamics and the measurements, we can proceed again by mapping the evolution of $\overline{\langle \mathbf{W}^{(n)} \rangle}$ to a reduced system with an effective Hamiltonian. If we focus again on the twofold replica $n=2$, we see that the action of the measurement operator on the reduced Hilbert space at each site is
\begin{equation}
\label{eq:measurementop}
\begin{split}
    &\mathcal{M} \ket{\mathbf{I}} = d\sum_m P_m^{\otimes 4} \ket{\mathbf{I}} = d\sum_m \ket{m}^{\otimes 4} \vcentcolon= \ket{\mathbf{O}}, \\
    &\mathcal{M} \ket{\mathbf{S}} = d\sum_m P_m^{\otimes 4} \ket{\mathbf{S}} = \ket{\mathbf{O}}, \\
    &\mathcal{M} \ket{\mathbf{O}} = d\sum_m P_m^{\otimes 4}\sum_n \ket{n}^{\otimes 4} = \ket{\mathbf{O}}.
\end{split}
\end{equation}

Therefore, we find that the new vector space $V_{\mathcal{M}}(S_2) = \operatorname{span}(\ket{\mathbf{I}},\ket{\mathbf{S}},\ket{\mathbf{O}})$ is closed under measurements. If we promote this to the reduced Hilbert space of the entire chain $V^{\otimes N}_{\mathcal{M}}(S_2)$, we find that this is also closed under the action of the Haar averaged unit cell between any pair of sites. To show this, we can consider the properties of the following linear combination
\begin{equation}
    \ket{\mathbf{X}} \vcentcolon= \ket{\mathbf{O}} - \frac{d}{d+1}(\ket{\mathbf{I}}+\ket{\mathbf{S}}) \in V_{\mathcal{M}}(S_2).
\end{equation}

It is straightforward to show that this becomes null under any contraction between a normal and a complex conjugate leg. Due to the rules of the Weingarten calculus, this means that such local states are preserved by averaged unit cells. This is summarised in the following equation
\begin{equation}
    \mathcal{T}_{ij}^{\mathcal{M}} \ket{\mathbf{X}}\otimes V_{\mathcal{M}}(S_2) \in \ket{\mathbf{X}}\otimes V_{\mathcal{M}}(S_2).
\end{equation}

In Appendix \ref{app:Xstate} we give an explicit representation of the new operator $\mathcal{T}_{ij}^{\mathcal{M}}$, acting on $V_{\mathcal{M}}(S_2)^{\otimes 2}$. We find that evolution in subspaces that contain $\ket{\mathbf{X}}$ states happen at a different rate $\Gamma$, independent of the rate of information propagation $\gamma$. This is defined by
\begin{equation}
\label{eq:Gamma}
    \Gamma = \frac{2d}{(d^2-1)^2} \operatorname{tr}(\operatorname{tr}_1(H)^2),
\end{equation}
and can be qualitatively understood as an energy cost associated with $\ket{\mathbf{X}}$ states. In App. \ref{app:micromacro} we derive a more explicit relation between the rates $\Gamma,\, \gamma$ and the microscopic Hamiltonian $H$.

The new state $\ket{\mathbf{X}}$, which appears after a measurement, ensures that we obtain the correct correlations between measurements performed consecutively at short time intervals on the same qudit. The timescale $1/\Gamma$ represents the time it takes a qudit to relax before we can obtain new information by measuring again in the same basis. For the rest of this work, we set $\Gamma \to \infty$, such that no measurement inertia can be observed. In App. \ref{app:Xstate}, we show that doing so is effectively equivalent to projecting out the $\ket{\mathbf{X}}$ state and working in the previous 2-dimensional reduced Hilbert space $V(S_2)$. The action of the measurements is also projected onto this subspace and can be expressed as
\begin{equation}
\label{eq:meas}
    \mathcal{M} = \frac{d}{d+1}(1+\sigma_x).
\end{equation}
It can be shown that this same operator is obtained in the reduced subspace if we consider instead measurements in random bases.

In the following, the measurements are distributed through the circuit according to an independent Poisson process for each site, at some uniform rate $f$ (in the natural time units of the continuous model).
The transfer matrix at time $t$ under both random dynamics and measurements is then given by an effective imaginary-time evolution $\mathcal{T}_{\rm eff}(t) = \exp (-t \mathcal{H}_{\rm eff})$, with $\mathcal{H}_{\rm eff}$ given by
\begin{equation}
    \mathcal{H}_{\rm eff} = \sum_{i=1}^{N-1} \mathcal{H}_{i,i+1} -f\sum_{i=1}^N \left(\mathcal{M}_i-1\right).
\end{equation}

From Eq. \ref{eq:entropy} and Eq. \ref{eq:ententropy} we see that we can express the second moment of the entanglement entropy of some sub-region $A$ at time $t$ using matrix elements of the transfer matrix 
\begin{equation}
\label{eq:S2A}
    \Tilde{S}_A^{(2)} = -\log \abs{\frac{\bra{\Psi_{A_{out}}}\mathcal{T}_{\rm eff}(t)\ket{\Psi_{A_{in}}}}{\bra{\mathbf{I}}^{\otimes N} \mathcal{T}_{\rm eff}(t)\ket{\Psi_{A_{in}}}}}.
\end{equation}

The denominator acts as a normalization factor, so using the expression above allows us to safely drop constant terms in the effective Hamiltonian.

We can perform a similar analysis for the case of multiple replicas. Using the results in App.~\ref{app:multiplereplicas} and the limits $d,\Gamma \to \infty$ we show that the effective Hamiltonian of the $n$'th replica theory is given by
\begin{equation}
\mathcal{H}^{(n)}_{\rm eff} = \frac{\gamma}{2}\sum_{i=1}^{N-1} D_{ij} - f\sum_{i=1}^{N} \mathcal{M}^{(n)},
\end{equation}
where $\mathcal{M}^{(n)}$ is the generalization of the operator in Eq.~\eqref{eq:meas} that acts as
\begin{equation}
\mathcal{M}^{(n)}\ket{\tau} = \sum_{\sigma \in S_n}\ket{\sigma},
\end{equation}
and $D$ is a diagonal two-site operator with entries given by
\begin{equation}
D_{\kappa\epsilon,\sigma\tau} = \delta_{\kappa\sigma}\delta_{\epsilon\tau} D(\sigma,\tau),
\end{equation}
with $D(\sigma,\tau)$ the bi-invariant metric on $S_n$ given by the Hamming distance between $\sigma$ and $\tau$, i.e. the number of elements that are not mapped onto themselves under $\tau^{-1}\sigma$. This form is manifestly consistent with the expected symmetry group $S_n \times S_n$. It is interesting to note that the $d\to \infty$ limit does not result in the fine tuned $S_{n!}$-symmetric Potts model observed in circuits with fully Haar random unit cells \cite{Bao2020}.

\section{Fermionic mapping}
\label{sec:fermionic}

If we take the limit of large local dimension $d$ and keep only the leading contributions, we obtain dynamics driven by
\begin{equation}
\label{eq:hamilteff}
    \mathcal{H}_{\rm eff} = -\frac{\gamma}{2}\left(\sum_{i=1}^{N-1}\sigma_z^{(i)}\sigma_z^{(i+1)}+g\sum_{i=1}^N \sigma_x^{(i)}\right),
\end{equation}
where $g = 2f/\gamma$. This is easily recognized as the transverse field Ising model (TFIM) in 1D, subject to open boundary conditions. It is well-known that this can be mapped to a system of non-interacting fermions using the Jordan-Wigner transformation \cite{Mbeng2020}. In this section, we will introduce the general formalism used to compute quantities of the form shown in Eq. \ref{eq:S2A}.

We start by constructing a set of non-local Majorana operators as
\begin{align}
        \gamma_i^{(1)} &= \sigma_z^{(i)}\prod_{j>i}\sigma_x^{(j)}, \\
        \gamma_i^{(2)} &= \sigma_y^{(i)}\prod_{j>i}\sigma_x^{(j)} = -i\sigma_z^{(i)}\prod_{j\geq i}\sigma_x^{(j)},
\end{align}
defined on all sites $i = 1,2 \dots N$ \cite{Kitaev2001}. These operators are Hermitian $(\gamma^{\mu})^\dagger = \gamma^{\mu}$ and obey the standard anticommutation relations
\begin{equation}
    \left\{\gamma^{\mu},\gamma^{\nu}\right\} = 2\delta^{\mu\nu},
\end{equation}
where the indices $\mu$, $\nu$ are understood to run over all $2N$ previously defined operators. From the definition, we get the additional relation
\begin{equation}
    \label{eq:sigmax}
    \gamma_i^{(1)}\gamma_i^{(2)} = -i\sigma_x^{(i)},
\end{equation}
such that the product of all Clifford operators is
\begin{equation}
\label{eq:chargeconj}
    \prod_{i=1}^{N} \gamma_i^{(1)}\gamma_i^{(2)} = (-i)^N\prod_{i=1}^N \sigma_x^{(i)}\coloneqq (-i)^N\mathcal{C}.
\end{equation}

This operator anti-commutes with all the Majorana fermions $\left\{\mathcal{C},\gamma^{\mu}\right\} = 0$ and it is a conserved quantity, since it commutes with the full Hamiltonian $\left[\mathcal{C},\mathcal{H}\right] = 0$. We can couple Majorana fermions living on adjacent sites into domain wall creation and annihilation operators
\begin{align}
\label{eq:fermops}
    a_i^{\dagger} &= \frac{1}{2}\left(\gamma_i^{(1)}-i\gamma_{i+1}^{(2)}\right), \\
    a_i &= \frac{1}{2}\left(\gamma_i^{(1)}+i\gamma_{i+1}^{(2)}\right),
\end{align}
for $i = 0,1 \dots N-1$, where we assume periodic boundary condition $N=0$. These obey the typical anti-commutation relations
\begin{equation}
\left\{a_{i}, a_{j}\right\}=0, \quad\left\{a_{i}^{\dagger}, a_{j}^{\dagger}\right\}=0, \quad\left\{a_{i}, a_{j}^{\dagger}\right\}=\delta_{ij}.
\end{equation}

A simple calculation shows that
\begin{equation}
a_i^{\dagger} a_i = \frac{1}{2}\left(1-\sigma_z^{(i)}\sigma_z^{(i+1)}\right),
\end{equation}
such that the number operator of the fermionic mode at some site $i \neq 0$ is a projector onto configurations that have a domain wall between sites $i$ and $i+1$. With this convention, the Hamiltonian becomes a quadratic form
\begin{equation}
    \mathcal{H}_{\rm eff} = \frac{\gamma}{2}\left[\sum_{i=1}^{N-1}a_i^{\dagger}a_i -g \sum_{i=1}^N\left(a_i^{\dagger}+a_i\right)\left(a_{i-1}-a_{i-1}^{\dagger}\right)\right].
    \label{eq:HamFermionic}
\end{equation}

This can be more succinctly expressed using the Bogoliubov-de Gennes notation
\begin{equation}
    \mathcal{H}_{\rm eff} = \frac{1}{2}\mathbf{a^{\dagger}}\mathcal{D}\mathbf{a},
    \label{eq:HamFermionGrand}
\end{equation}
where $\mathbf{a}= (a_0,\,a_1,\,\dots\, ,\, a_{N-1},\,a_0^{\dagger},\, a_1^{\dagger},\, \dots \, ,\, a_{N-1}^{\dagger})^T $. The matrix $\mathcal{D}$ is called the grand-dynamical matrix and it obeys the particle-hole symmetry equation 
\begin{equation}
\label{eq:parthole}
    \eta\mathcal{D}^{T}\eta = -\mathcal{D},\,
    \mathrm{where}\;
    \eta = 
    \begin{bmatrix}
    0 & I \\
    I & 0
    \end{bmatrix}.
\end{equation}.

The exponentials of such Hamiltonians are most easily treated using the algebra of fermionic Gaussian states, as worked out in Ref. \cite{Fagotti2010}. We will briefly outline some of the results relevant to our calculation. It is convenient to define a Gaussian state through its generating quadratic form as
\begin{equation}
\label{eq:rho}
    \rho[W] = \frac{1}{Z(W)}\exp\left(\frac{1}{2}\textbf{a}^{\dagger}W\textbf{a}\right),
\end{equation}
with a normalisation constant $Z(W)$ chosen such that $\Tr \rho[W] = 1$. By Wick's theorem, such many-body states are fully characterized by their two-body correlation matrix, defined as
\begin{equation}
\label{eq:corrmatrix}
    \Gamma_{\mu\nu} = 2\Tr \left(\rho[W] \textbf{a}^{\dagger}_\mu\textbf{a}_\nu\right) - \delta_{\mu\nu}.
\end{equation}

The correlation matrix is related to the generator of the quadratic form through the useful relations
\begin{equation}
    \Gamma = \tanh\left(\frac{W}{2}\right),\, e^W = \frac{1+\Gamma}{1-\Gamma},
\end{equation}
where it is assumed that $1-\Gamma$ is invertible. Using a special case of the Baker-Campbell-Hausdorff formula, it is shown that fermionic Gaussian states are closed under multiplication and we have
\begin{equation}
    \rho[\Omega] = \frac{Z(W)Z(W')}{Z(\Omega)}\rho[W] \rho[W'],
\end{equation}
with the new generating matrix $\Omega$ given by
\begin{equation}
    \Omega = \log(\exp(W)\exp(W')).
\end{equation}

If we denote the correlation matrix of $\Omega$ by $\Gamma \times \Gamma'$, with $\Gamma$ and $\Gamma'$ the correlation matrices of $W$ and $W'$ respectively, the following formula is proven in Ref. \cite{Fagotti2010}
\begin{equation}
\Gamma \times \Gamma'=1-\left(1-\Gamma^{\prime}\right) \frac{1}{1+\Gamma \Gamma'}(1-\Gamma).
\end{equation}

Inner products can be easily computed using the following trace formula
\begin{equation}
\label{eq:overlaps}
    \{\Gamma,\Gamma'\} = \Tr \left(\rho[W]\rho[W']\right) = \pm \sqrt{\abs{\det\frac{1+\Gamma \Gamma'}{2}}},
\end{equation}
where the ambiguity of the sign is in general a complex issue, but this will not be a problem for our purposes.

\section{Dynamics of purification}
\label{sec:QuantitativeResults}

In the preceding sections, we developed a formalism that allows us to study entanglement dynamics in our continuous-time random quantum circuit models. Here, we focus specifically on purification dynamics in these models. Namely, starting from an initial mixed state, we are interested in how fast the state of the system is purified by measurements. We will be particularly interested in the purification transition that occurs as a function of measurement frequency $f$ \cite{Gullans2020}, which is thought to be concomitant with the measurement-induced entanglement transition separating area- and volume-law phases \cite{Li2018,Skinner2019,Chan2019,Li2019,Choi2020}. Thanks to the exact solvability of our model in the $d \rightarrow \infty$ limit, we are able to compute analytical expressions for the order parameters of this dynamical phase transition, and infer the key critical exponents.

\subsection{Setup and phase diagram}

The setup we study is as in Ref.~\cite{Gullans2020}: The system is initialized in a maximally mixed state, which is represented in the above formalism by the input state $\ket{\mathbf{I}}^{\otimes N}$. After some evolution time $t$, the purity of the state of the system will have increased from its initial value due to the measurements. As explained previously, we will use the quantity \eqref{eq:ententropy} as a measure of the the typical entropy of the ensemble of states. Since we are looking at the purity of the entire state after a time $t$, the set $A$ that appears in Eq.~\eqref{eq:S2A} will contain all of the output qubits. Accordingly, we can express the quantity in question in terms of the transfer matrix $\mathcal{T}_{\rm eff}(t)$
\begin{equation}
    \tilde{S}^{(2)}(t) = -\log \abs{\frac{\bra{\mathbf{S}}^{\otimes N} \mathcal{T}_{\rm eff}(t)\ket{\mathbf{I}}^{\otimes N}}{\bra{\mathbf{I}}^{\otimes N} \mathcal{T}_{\rm eff}(t)\ket{\mathbf{I}}^{\otimes N}}}.
    \label{eq:SLogRatio}
\end{equation}

The purification transition that occurs in our model is associated with a quantum phase transition in the effective Hamiltonian $\mathcal{H}_{\rm eff}$, which generates the time evolution operator $\mathcal{T}_{\rm eff}(t)$. Based on the phase diagram of the TFIM, we can deduce that such a transition must occur at the critical measurement rate $g_c = 1$, i.e.~$f_c = \gamma/2$. In the spin basis \eqref{eq:hamilteff}, the two phases correspond to the $\mathbb{Z}_2$ symmetric phase under the symmetry $\mathcal{C} = \prod_{i=1}^N \sigma_x^{(i)}$ for $g > 1$, and a spontaneous symmetry-broken phase for $g < 1$.

For the problem in hand, the relevant order parameter that we use to distinguish these phases is not a correlation function, as is usually the case, but rather the many-body overlap appearing inside the logarithm in Eq.~\eqref{eq:SLogRatio}. To provide intuition into how this quantity behaves either side of the transition, we can reformulate our expression for $\tilde{S}^{(2)}(t)$ as follows. Since $\ket{\mathbf{S}}^{\otimes N} = \mathcal{C} \ket{\mathbf{I}}^{\otimes N}$, the above fraction becomes equal to the expectation value of $\mathcal{C}$ in the state $\ket{\boldsymbol\Psi (t)} = \mathcal{T}_{\rm eff}^{\frac{1}{2}}\ket{\mathbf{I}}^{\otimes N}$, namely
\begin{equation}
    \Tilde{S}^{(2)}(t) = - \log \abs{\langle \mathcal{C} \rangle_{\boldsymbol\tilde{\Psi}(t)}},
\end{equation}
where $\ket{\boldsymbol\tilde{\Psi}(t)} \coloneqq \ket{\boldsymbol\Psi (t)} / \sqrt{ \braket{\boldsymbol\Psi (t) | \boldsymbol\Psi (t)}}$ is the wavefunction after imaginary time evolution under $\mathcal{H}_{\rm eff}/2$, appropriately normalized.

If the measurement rate is sufficiently high such that the Hamiltonian \eqref{eq:hamilteff} is in a symmetry-unbroken phase, then the ground state is non-degenerate and thus $\ket{\boldsymbol\tilde{\Psi}(t)}$ inherits the symmetry of the Hamiltonian. Since the Hamiltonian is also gapped, we see that the (accordingly normalized) state $\ket{\boldsymbol\Psi (t)} = \exp(-t\mathcal{H}_{\rm eff}/2)\ket{\mathbf{I}}^{\otimes N}$ converges to the ground state exponentially quickly. The ground state must be an eigenstate of $\mathcal{C}$, whose eigenvalues are $\pm 1$, so we can then conclude that $\abs{\langle \mathcal{C} \rangle_{\boldsymbol\Psi(t)}} \to 1$ exponentially quickly as $t \to \infty$, and hence $\Tilde{S}^{(2)} \to 0$ at a rate independent of the system size, as expected in this regime. When $g<1$ the symmetry is spontaneously broken. In this case, the ground eigenspace is doubly degenerate in the thermodynamic limit $N \rightarrow \infty$, and the effect of the transfer matrix at long times is to project onto this subspace. The projected state may no longer be an eigenstate of $\mathcal{C}$, so we can have a non-zero residual entropy. As we will see, this residual entropy is extensive, with a $\log (N)$ correction [Eq.~\eqref{eq:EntropyScaling}].\\

While this picture allows us to understand the transition at a qualitative level, to obtain an analytic expression for the residual entropy we will instead use the fermionic mapping detailed in the previous section. We will find it convenient to work with states of definite fermion parity, and hence we define the density matrices $\rho_{\pm} = \ket{\pm}\bra{\pm}$, where $\ket{\pm} \coloneqq (\ket{\mathbf{I}}^{\otimes N} \pm \ket{\mathbf{S}}^{\otimes N})/\sqrt{2}$, which are eigenstates of $\mathcal{C}$. We can then write
\begin{equation}
    \Tilde{S}^{(2)} = \log\abs{\frac{1+\Theta}{1-\Theta}},
    \label{eq:SLogTheta}
\end{equation}
where the parameter $\Theta$ is defined by
\begin{equation}
\label{eq:ThetaRho}
    \Theta = \frac{\Tr\left(e^{-t \mathcal{H}_{\rm eff}}\rho_{+}\right)}{\Tr\left(e^{-t \mathcal{H}_{\rm eff}}\rho_{-}\right)},
\end{equation}
We note that Eqs.~(\ref{eq:SLogTheta}, \ref{eq:ThetaRho}) are quite general, and could be applied even if we didn't take the $d \rightarrow \infty$ limit.

Because the Hamiltonian \eqref{eq:HamFermionic} is a fermion bilinear, the exponential $e^{-t \mathcal{H}_{\rm eff}}$ can be written in the form of Eq.~\eqref{eq:rho}, with the grand dynamical matrix $\mathcal{D}$ in place of $W$. Hence we can define correlation matrices $\Gamma[-t \mathcal{D}]$ that correspond to this fermionic state, according to Eq.~\eqref{eq:corrmatrix}. The states $\rho_{\pm}$ are also Gaussian fermionic states, and hence can be characterized through their correlation matrices. These have the simple diagonal form $\Gamma_{GS} = \operatorname{diag}(-1,\,-1,\, \dots ,\, -1,\, 1,\,1,\,\dots,\,1)$ and $\Gamma_{E} = \operatorname{diag}(1,\,-1,\, \dots ,\, -1,\,-1,\,1,\,\dots,\,1)$, with $\pm1$ each appearing $N$ times. Then, using Eq.~\eqref{eq:overlaps} we obtain
\begin{equation}
\label{eq:Theta}
    \Theta = \frac{\{\Gamma[-t\mathcal{D}],\Gamma_{+}\}}{\{\Gamma[-t\mathcal{D}],\Gamma_{-}\}}.
\end{equation}
This expression for $\Theta$, which determines the \RE entropy via Eq.~\eqref{eq:SLogTheta}, will help us study the purification transition at a quantitative level.\\

The above considerations help us anticipate the existence of two distinct dynamical phases, consistent with previous work on purification dynamics in discrete time random circuit models, which we refer to as `mixed' ($g < 1$) and `purifying' phases ($g > 1$), following Ref.~\cite{Gullans2020}. In the following, we derive analytical expressions for the time dependence of $\Theta$, which in turn determines the \RE entropy $\tilde{S}^{(2)}(t)$ via Eq.~\eqref{eq:SLogTheta}.  We will use these expressions later to understand the nature of the two phases and the transition between them at a quantitative level.

\subsection{Expressions for $\Theta(t)$}

While we have so far left the boundary conditions unspecified, in computing $\Theta(t)$ we will consider open and periodic boundary conditions separately in our calculations. The conventional timescale $\gamma = 1$ is employed throughout this chapter.

\subsubsection{Periodic boundary conditions}

We start by considering periodic boundary conditions, which can be realised by introducing additional random unitary gates that act between sites 1 and $N$ in the original circuit model. In this case, a standard calculation shows that the Jordan-Wigner-transformed Hamiltonian \eqref{eq:HamFermionic} acquires an additional term which imposes either periodic or antiperiodic boundary conditions depending on the fermion parity sector one works in (see, e.g.~Ref.~\cite{Mbeng2020}). Taking $N$ to be even from hereon for simplicity, the even (odd) parity sector features antiperiodic (periodic) boundary conditions. These are sometimes referred to as Ramond and Neveu-Schwarz sectors, respectively.

Thanks to the translation invariance of the system, the single particle Hamiltonian $\mathcal{D}$ can be block diagonalized using momentum eigenstates, whose wavevector is quantized to $k_l = l\pi/N$, with $l \in \{1, \ldots, N-1\}$. In the even parity sector $l$ must be odd to be compatible with the boundary conditions, and likewise \textit{vice-versa}. Recognizing that the states $\ket{\pm}$ are the ground states of the Hamiltonian in the $g \rightarrow 0$ limit in each parity sector, we can express $\Theta$ as a ratio of products over $k_l$ modes, with even $l$ in the numerator and odd $l$ in the denominator. In Appendix \ref{app:pbc}, we show that
\begin{equation}
\Theta = e^{t} \frac{\prod_{n=1}^{N/2 - 1}\theta(k_{2n}, t)}{\prod_{n=1}^{N/2}\theta(k_{2n-1}, t)},
\label{eq:ThetaProductPBC}
\end{equation}
where
\begin{equation}
    \theta(k, t) = \cosh(\lambda_k t)\left[1 + \tanh(\lambda_k t)\frac{1-g \cos k}{\lambda_k}\right],
\end{equation}
and the energy eigenvalues are given by
\begin{align}
    \lambda_k = \sqrt{1+g^2 - 2g \cos k}.
    \label{eq:Dispersion}
\end{align}
The factor $e^{t}$ in Eq.~\eqref{eq:ThetaProductPBC} accounts for the modes $k = 0$, $k = \pi$, which are only present in the odd parity sector.

\subsubsection{Open boundary conditions}

We can also consider open boundary conditions (OBCs), where the Hamiltonian $\mathcal{H}_{\rm eff}$ no longer features the term connecting sites 1 and $N$. While the change of boundary condition makes little difference in the purifying phase, we will find that in the mixed phase, quantitative differences between OBCs and PBCs can be seen in the behaviour of the \RE entropy. As such, will focus mainly on mixed phase $g < 1$, although of the following holds true throughout the phase diagram.

With OBCs, the JW-transformed does not contain a term that manifestly depends on the fermion parity sector. This leaves us with the problem of diagonalizing the single-particle matrix $\mathcal{D}$, which is now the same in both parity sectors. Since the system is no longer translation invariant, we cannot treat momentum eigenmodes separately as we did before. Instead, we must explicitly compute the correlation matrices $\Gamma[-t\mathcal{D}]$ and $\Gamma_{\pm}$, and use the more general expression \eqref{eq:Theta}. To do so, we calculate the single-particle eigenstates, which form the columns of a real orthogonal eigenvector matrix $\mathcal{O}$. In terms of these, we obtain a spectral decomposition of the grand dynamical matrix $\mathcal{D} = \mathcal{O} \Lambda \mathcal{O}^{-1}$. The eigenvalues $\Lambda$ come in pairs due to the particle-hole symmetry \eqref{eq:parthole}, which we arrange as $\Lambda = \operatorname{diag}(\lambda_0,\,\lambda_1,\,\lambda_2,\, \ldots \lambda_{N-1}, -\lambda_0,\,-\lambda_1,\,-\lambda_2,\,\ldots -\lambda_{N-1})$ with $\lambda_i > 0$ and in non-decreasing order. The corresponding pairs of eigenvectors are also related through $\ket{-\lambda_i} = \eta \ket{\lambda_i}$, with $\eta$ defined in Eq.~\eqref{eq:parthole}. In terms of these eigenvalues, the correlation matrix $\Gamma[-t\mathcal{D}]$ becomes
\begin{align}
    \Gamma[-t\mathcal{D}] = \mathcal{O} \tanh (\frac{-t\Lambda}{2})  \mathcal{O}^{-1},
\end{align}
which can be substituted directly into Eq.~\eqref{eq:Theta}. In Appendix \ref{app:prodrep}, we show that the single particle eigenstates take the form of sinusoids, whose wavevectors $k_l$ can be found as the solutions to the equation
\begin{equation}
    \tan N k_l = \frac{g\sin k_l}{1-g \cos k_l},
    \label{eq:OBCWavevector}
\end{equation}
lying in the interval $[0, \pi )$ and labeled in increasing order. In terms of these wavevectors, the eigenvalues $\lambda_l$ themselves again follow the well-known dispersion for the TFIM, Eq.~\eqref{eq:Dispersion}. In the mixed phase, there is also a single imaginary solution to the above, which we label $k_0 = \iu K$, corresponding to a Majorana edge mode localized at the two boundaries of the chain. The energy of this edge mode $\lambda_0$ can be shown to be exponentially small in the system size $N$, and this energy is associated with a long timescale $\lambda_0^{-1}$ which we will show is responsible for the slow decay of purity in this phase.

\subsection{Behaviour of the \RE entropy}

With the above expressions in place, we are now ready to study the behaviour of the \RE entropy in the two phases, as well as the critical point which separates them.

\subsubsection{Mixed phase---periodic boundary conditions}

In the mixed phase, the \RE entropy shows quantitative differences depending on the boundary conditions, and thus we will consider both PBCs and OBCs in the regime $g < 1$, starting with the former. Beginning with Eq.~\eqref{eq:ThetaProductPBC}, we can use complex integration methods to transform the alternating product over even and odd $k$ modes into an infinite product
\begin{equation}
\label{eq:ThetaInfiniteProduct}
    \Theta = \prod_{q=0}^\infty \tanh \frac{Nx_q}{2},
\end{equation}
with $x_q$'s found as the solutions of the equation
\begin{align}
\label{eq:xqeq}
    &t\sqrt{2g\cosh x-1-g^2}+\phi(x) = \pi(q+\frac{1}{2}), \\
    &\tan \phi(x) = \frac{g\cosh x - 1}{\sqrt{2g\cosh x-1-g^2}},
\end{align}
in the interval $(K,\infty)$, where $K = -\log g$ is the point in the complex plane where the dispersion function $\lambda(\iu K)$ has a zero. This form makes it manifestly clear that $\Theta < 1$. The details of this calculation are given in App. \ref{app:pbc}. Except at criticality, for sufficiently large $N$ we can have $NK \gg 1$, which in turn implies $Nx_q \gg1$ for all $q$. In this case, by virtue of the approximation $\log \tanh(y) \approx -2e^{-2y}$ for $y \gg 1$, we can approximate $\log \Theta$ by an integral
\begin{equation}
    \log \Theta \approx -2\int_K^\infty \frac{dq}{dx} e^{-Nx},
\end{equation}
where $dq/dx$ is the density of solutions and can be found by differentiating Eq.~\eqref{eq:xqeq}. The result to highest order in powers of $N$ is
\begin{equation}
    \log \Theta(t) = -\sqrt{\frac{1-g^2}{\pi N}}e^{-NK}\left(t + \frac{2}{1-g^2}\right).
\end{equation}
To relate this expression to the \RE entropy \eqref{eq:ententropy}, we focus on the regime where $t$ scales no faster than polynomial in $N$, such that the small factor $e^{-NK}$ dominates, making $-\log \Theta(t)$ itself small. Then, if we make the approximations $1-\Theta \approx -\log \Theta$ and $1+\Theta \approx 2$, we can deduce the following form for the \RE entropy, valid for a broad window of times $e^{NK} \gg t \gg 2/(1-g^2)$ (restoring the original units of time)
\begin{align}
    \tilde{S}^{(2)}(t) = N\log|g| - \log \frac{t}{\sqrt{N}} + \frac{1}{2}\log \frac{4\pi}{1-g^2} + o(1)
    \label{eq:SLogDecreasePBC}
\end{align}
where the term $o(1)$ represents terms that tend to zero in the limit of large $t$ or $N$. We see that the entropy decreases very slowly in time in this window, which is a defining feature of the mixed phase.

Finally, for times $t$ that scale exponentially with system size $\gamma t \gtrsim e^{NK}$, such that $|\log \Theta(t)| \gg 1$, we find that the entropy decays as
\begin{equation}
\label{eq:entropylatePBC}
    \Tilde{S}^{(2)} \approx 2\Theta(t) \approx 2\exp\left(-\gamma t\sqrt{\frac{1-g^2}{\pi N}} e^{-NK} \right)
\end{equation}

\subsubsection{Mixed phase---open boundary conditions}

We now wish to compute the same quantity with open boundary conditions in the mixed phase $g < 1$. In particular, we are interested in the regime during which the entropy decays very slowly. As such, we can separate out the bulk single-particle eigenstates, whose energies lie above the bulk gap $\Delta = (1-g)$, from the Majorana edge mode, which is exponentially small in $N$. In particular, as long as one is not too close to criticality $(1-g) \gg 1/N$, the Majorana eigenvalue can be approximated as
\begin{equation}
    \lambda_0 \approx (1-g^2)e^{-NK}
\end{equation}
This indicates that there is a regime of times $\Delta^{-1} \ll t \ll \lambda_0^{-1}$ during which the transient bulk modes have decayed away $\exp(-t\lambda_i) \approx 0$, while the Majorana mode has not decayed. 
The approximate correlation matrix in this regime will then take the form
\begin{align}
\label{eq:Gammaapprox}
    \Gamma&[-t\mathcal{D}] = \mathcal{O} \tanh (-\frac{t\mathcal{E}}{2})  \mathcal{O}^{-1} \nonumber\\
    &\approx \mathcal{O} \begin{pmatrix}
    -\tanh \frac{t\lambda_0}{2} & 0 & 0 &  0 \\
    0 & -I_{N-1} & 0 &  0 \\
    0 & 0 & \tanh \frac{t\lambda_0}{2} &  0 \\
    0 & 0 & 0 & I_{N-1}
    \end{pmatrix} \mathcal{O}^{-1}.
\end{align}
\begin{figure}[t!]
    \includegraphics[width=0.47\textwidth]{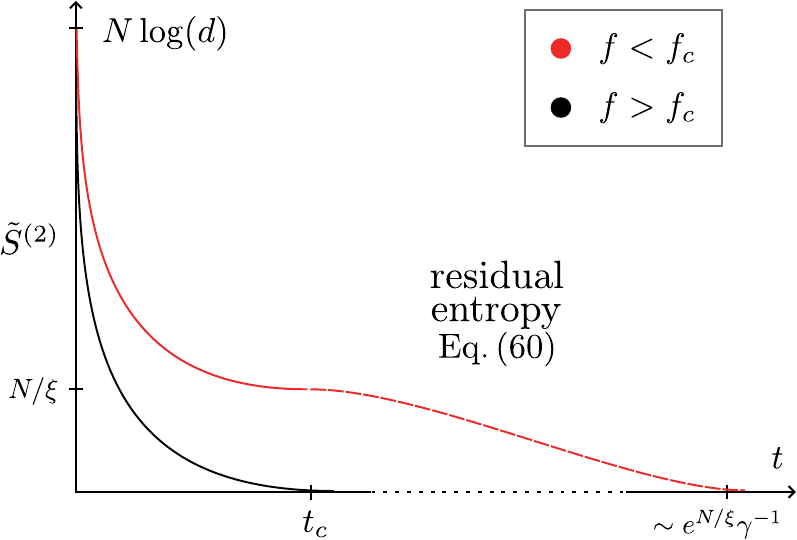}
    \caption{Qualitative plot showing the behaviour of the system entropy on different timescales in the mixed(red) and purifying(black) phase. In the mixing phase, the entropy undergoes a period of very slow logarithmic decay up to exponentially long times in the system size.}
    \label{fig:resentropy}
\end{figure}

In App. \ref{app:prodrep} we show that the form of the eigenvectors $\mathcal{O}$ can be found explicitly and the parameter $\Theta$ can be expressed using Vandermonde determinants. The factorization of the latter is known, leading to the exact final expression
\begin{equation}
\label{eq:Thetaprod}
    \Theta(t) = e^{-t\lambda_0}\frac{\tanh \frac{NK}{2}\prod_{l \text{ odd}} (\cosh K - \cos k_l)}{\sinh K \prod_{l \text{ even}} (\cosh K - \cos k_l)},
\end{equation}
where $k_l$ for $1 \leq l \leq N-1$ are the wavevectors of the bulk modes, defined in Eq.~\eqref{eq:OBCWavevector}, and $K$ is the spatial decay rate of the edge mode. As with the analogous expression for periodic boundary conditions \eqref{eq:ThetaProductPBC}, this product can be evaluated with the help of complex integration techniques, which we describe in Appendix \ref{app:largeN}. We find that the large $N$ asymptotic expression of $\Theta$ takes the form
\begin{equation}
    \log \Theta(t) = 2e^{-NK}\left(\sqrt{\frac{N}{\pi}}\sqrt{1-g^2}-1-\frac{1-g^2}{2}t\right)
\label{eq:EntropyScaling}
\end{equation}
Again, we focus on the regime where $t$ scales polynomially with $N$, in which case the right hand side of the above is small. Moreover, noting that $\Theta$ should be no greater than unity, we find that the above expression should only be trusted in the regime
\begin{equation}
    t \gtrsim t_c = 2\sqrt{\frac{N}{(1-g^2)\pi}},
\end{equation}
We view the above constraint as a condition for validity of the approximation \eqref{eq:Gammaapprox} made earlier. Then, in this regime we can make the same series of approximations as before to relate $\Theta(t)$ to the \RE entropy. Thus, for $t_c \lesssim t \ll e^{NK}$, we find
\begin{align}
    \label{eq:entropyearly}
    \tilde{S}^{(2)}(t) = N\log|g| - \log t + \log \frac{2}{1-g^2} + o(1)
\end{align}
At very long times, when $t$ scales exponentially with $N$ such that $\abs{\log \Theta(t)}\gg 1$, we find that the entropy decays as
\begin{equation}
\label{eq:entropylate}
    \Tilde{S}^{(2)}  \approx 2\exp\left(-(1-g^2) te^{-NK}\right)
\end{equation}
Together, Eqs.~(\ref{eq:entropyearly}, \ref{eq:entropylate}) characterize the salient features of purification dynamics in our model in the mixed phase sufficiently far from criticality ($e^{-NK} \ll 1$).

\subsubsection{Purifying phase}

The purifying case corresponds to the regime $g > 1$, where measurements occur so often that they overcome the scrambling and an initially mixed state quickly becomes pure. Looking at the spectrum of the single-particle matrix $\mathcal{D}$, one finds that all eigenvalues are at least as large as the bulk gap $\Delta = (g-1)$, which sets a timescale $t \gtrsim (g-1)^{-1}$ after which the correlation matrix $\Gamma[-t\mathcal{D}]$ will have converged close to its $t \rightarrow \infty$ limit. 

Using Eq. \ref{eq:ThetaInfiniteProduct} we see that there is exactly one root $x_0$ in the interval $0 < x_0 < \log g$. For large enough $N$ we have that $N\log g \gg 1$, so for the other roots $\tanh Nx_q/2 \approx 1$. This means that for $t \sim \text{poly}(N)$ we can make the approximation
\begin{equation}
    \Tilde{S}^{(2)}(t) \approx Nx_0,
\end{equation}
so $x_0$ is the entropy density in the chain at time $t$. The equation determining $x_0$ can be written as
\begin{equation}
    \tanh\left(t\sqrt{1+g^2-2g\cosh x_0}\right) = \frac{\sqrt{1+g^2-2g\cosh x_0}}{g\cosh x_0 - 1}.
\end{equation}

Taking $t \gg 1$ we see that the solution is approximately
\begin{equation}
    \Tilde{S}^{(2)}(t)/N \approx x_0 \approx \frac{2(g-1)}{g}e^{-t(g-1)},
\end{equation}
and follows the expected decay rate set by the spectral gap $\Delta$. Since the system is disordered in this regime, the result is expected to hold in the thermodynamic limit, irrespective of the boundary conditions.

\subsubsection{Critical point}

We now address dynamics at the critical point $g = 1$. While we are no longer able to reliably make an approximation of the kind \eqref{eq:Gammaapprox} in the open boundary condition case, we find that the purification dynamics for periodic boundary conditions is amenable to analytical treatment in this regime. In particular, Eq.~\eqref{eq:ThetaInfiniteProduct} continues to hold at $g = 1$, with the the equation for the roots $x_q$ now given as solutions to the equation
\begin{equation}
    2t\sinh \frac{x}{2} +\arctan \sinh \frac{x}{2} = \pi(q+\frac{1}{2}).
\end{equation}
This equation is still transcendental, making it difficult to find a universal expression for its solutions. However, we can study the three regimes $t \ll 1$, $1 \ll t \ll N$ and $t \gg N$ separately. 

\textit{Early times $t \ll 1$.---}In the initial time frame $t \ll 1$ it can be shown that the entropy is approximately given by
\begin{equation}
    \tilde{S}^{(2)} \approx -N\log \frac{t}{2}.
    \label{eq:EntropyCriticalAlgebraic}
\end{equation}
The logarithmic divergence at the origin has a simple intuitive explanation: since the averaging is performed over the purities rather than the entropies and we work in a $d\to \infty$ system, the only scenarios that contribute to the average purity at early times are those where the entire chain is measured. If this happens, the system is immediately purified, since we can neglect the unitary evolution at $t \ll 1$. The entropy is then simply the logarithm of the probability that all qudits are measured within the time $t$, which is $p \approx (ft)^N = (gt/2)^N$. This intuitive picture matches the exact answer we found above at criticality, and is expected to hold for all values of $g$ in both the periodic and open boundary conditions.

\textit{Intermediate times $1 \ll t \ll N$.---}In this regime we can assume $Nx_q \gg 1$ but $x_q \ll 1$ for all solutions $x_q$ that contribute meaningfully to the value of $\Theta$. Using these approximations we find that the $x_q$ are equally spaced and the formula for $\log \Theta$ is calculated as a geometric sum. The final expression for the entropy is
\begin{equation}
    \tilde{S}^{(2)} \approx \frac{N\pi}{2t+1}.
\end{equation}

The algebraic relationship $\tilde{S}^{(2)} \propto t^{-1}$ is an important feature and only occurs exactly at criticality.

\textit{Late times $t \gg N$.---}Finally, in the long time limit, we see that we can approximate $\Theta$ by
\begin{equation}
    \Theta \approx \frac{(e^{-\frac{N \pi}{2t}},e^{-\frac{N\pi}{t}})_\infty}{(-e^{-\frac{N \pi}{2t}},e^{-\frac{N\pi}{t}})_\infty}  \approx \sqrt{2}e^{-\frac{\pi t}{4N}},
\end{equation}
where $(a,q)_\infty$ is the q-Pochammer symbol. This leads to an exponential decay of the entropy
\begin{equation}
    \tilde{S}^{(2)} \approx 2\sqrt{2} e^{- \frac{\pi t}{4N}}.
\end{equation}

\subsection{Comparison to result from field theory}

Having derived expressions for the time-dependence of the R{\'e}nyi entropy of our model of hybrid quantum dynamics, it is instructive to compare our findings to the approach introduced in Ref.~\cite{Li2021}. There, the authors invoke an effective field theory known as capillary-wave theory, which was first developed to model the dynamics of domain walls in the low-temperature phase of the Ising model. The correspondence between the two is rooted in the mapping between discrete-time hybrid quantum circuits and two-dimensional ferromagnets, see e.g. Refs.~\cite{Bao2019, Skinner2019}. The parameters of the theory are a phenomenological surface tension $\sigma$ and inverse temperature $\beta$, and once these are fixed, it is possible to find approximations for the time-dependence of the \RE entropy starting from a mixed initial state in the associated discrete-time monitored quantum circuit model.

Upon comparing their expression to our results, we find that the same universal features hold. In particular, for both cases, there is a marked regime of times $t \sim \text{poly}(N)$ in the mixed phase during which the entropy decays as an extensive constant with a $-\log t$ contribution. The sensitivity to boundary conditions we see [Eq.~\eqref{eq:SLogDecreasePBC} vs.~\eqref{eq:entropyearly}] can also be understood in the capillary-wave picture as a consequence of the difference in configurational entropies of the endpoints of a domain wall for periodic versus open boundary conditions. Moreover, by looking at the prefactor of the term proportional to $N$, we can relate the microscopic parameters of our model to the phenomenological parameters of the field theory; in particular, we can fix the code rate $\beta \sigma = K = -\log g$, which vanishes non-analytically at the transition $g = 1$.

\section{Discussion}
\label{sec:Discussion}

Our work introduces a class of random unitary circuits following a brickwork geometry, where each unit cell performs an infinitesimally small unitary transformation. We show that the limiting case of the construction above leads to a continuous stochastic process through the many-body Hilbert space. We show that the non-equilibrium behaviour of statistical averages of a large class of operator-space entanglement measures (the \RE entropies) of this dynamical process can be obtained as equilibrium partition functions in an effective quantum spin system, governed by a universal, time-independent Hamiltonian. The construction relies on an initial microscopic Hamiltonian describing local interactions, but we prove that this only enters the effective quantum information dynamics by setting the overall timescale.

We only perform a thorough investigation of the second \RE entropy, where the effective theory is the spin-$1/2$ ferromagnetic TFIM, with an integrability breaking term that becomes quadratically small in the local dimension $d$. The ground state becomes degenerate in the thermodynamic limit and it is ferromagnetically ordered. Taking a phenomenological perspective, the two types of stable ordering roughly correspond to the measuring agent having full knowledge or no knowledge about the state of the system. The lowest energy excitations are topological domain walls, and roughly represent the geometric boundaries of our knowledge. We show that local measurements can also be studied within the same framework by adding an extra state to the spins of the effective system. When the local tumbling rate of the microscopic Hamiltonian is sufficiently strong, this extra state is adiabatically eliminated, and the effect of measurements is to introduce a transverse magnetic field whose strength is proportional to the measurement frequency. When this exceeds a critical threshold, the system undergoes an Ising-type phase transition into a disordered phase. This is recognized as the purification transition observed in numerical studies of similar models \cite{Gullans2020,Gopalakrishnan2021}. The signature of the transition is a logarithmically decaying residual uncertainty in the state of the system after a purification procedure using uncorrelated local measurements, which is present only if the system is in the ordered phase.

We identify the order parameter corresponding to the residual second \RE entropy in the effective model and prove exact product expansion formulae that can be used to calculate it in both open and closed boundary conditions. Complex integration techniques are used to find thermodynamic limit approximations on various timescales, and we see that their scaling agrees with field theoretic arguments. The method is not restricted to the residual entropy of the whole chain, and could be adapted to calculations of other second \RE entropies. Universal characteristics of the transition such as the critical exponents must be the same as for the effective 1D quantum Ising theory. The transition in the von Neumann entropy requires higher replica analysis and may be of a different universality class, but we expect a qualitatively similar behaviour away from criticality. We begin the investigation of higher replica calculations by proving a formula for the matrix elements of the effective Hamiltonian in App. \ref{app:multiplereplicas}. In contrast with similar models studied in literature, our circuits are not expected to lead to a percolation transition of the von Neumann entropy, even in the $d\to \infty$ limit. This is because the small gate action limit $\Delta t \to 0$ is taken first, making the Hartley entropy $S_0$ undefined for any $d$. 

To simplify our calculations, we have set the local tumbling rate $\Gamma$ to infinity, but it may be interesting to investigate how it affects the transition. This introduces measurement inertia, wherein less information is gained by consecutively measuring the same qudit at intervals less than $\sim 1/\Gamma$. If the measurement frequency grows beyond this, the qudits become effectively Zeno-locked. To the best of our knowledge, the growth of entanglement in this regime has not been previously investigated.

\begin{acknowledgements}
    This work was supported in part by EPSRC grant EP/S020527/1. S.L. acknowledges support from UCL’s Graduate Research Scholarship and Overseas Research Scholarship. M.~M.~acknowledges support from Trinity College, Cambridge.
\end{acknowledgements}

\appendix

\section{Diagrammatic calculations of matrix elements}
\label{app:diagrammatic}

In this appendix, we show how diagrammatic manipulations can be used to find the effective transfer matrix in Eq. \ref{eq:transfunitcell}. Our goal is to find how the unit cell defined in Sec. \ref{sec:unitmodel} can be understood as a linear map on $V(S_2)$. If we denote by $V^\star(S_k)$ the dual space of $V(S_k)$ then we can define the dual element $\sigma^\star$ of $\sigma$ by the following expression

\begin{equation}
    \bra{\sigma^\star}\ket{\tau} = \sigma^\star(\tau) =\delta_{\sigma,\tau}.
\end{equation}

It is then easy to check that the following expansion holds for all $\sigma$

\begin{equation}
    \ket{\sigma} = \sum_{\tau \in S_k}\bra{\tau^\star}\ket{\sigma} \ket{\tau},
\end{equation}
so we have a resolution of identity

\begin{equation}
    1 = \sum_{\sigma \in S_k} \ket{\sigma}\bra{\sigma^\star}.
\end{equation}

The dual element can be expanded on a basis formed from Hermitian conjugates of the original $V(S_k)$ basis, with coefficients given by the Weingarten function

\begin{equation}
    \bra{\sigma^\star} = \sum_{\tau \in S_k} \mathcal{W}g(\sigma\tau^{-1})\bra{\tau}.
\end{equation}

We can compute a matrix associated to the unit cell by considering contractions of its inputs and outputs with elements of $S_k$. We denote the result of this diagrammatic contraction by

\begin{figure*}[t!]
     \centering
     \begin{subfigure}[t]{0.45\textwidth}
         \centering
         \includegraphics[width=\textwidth]{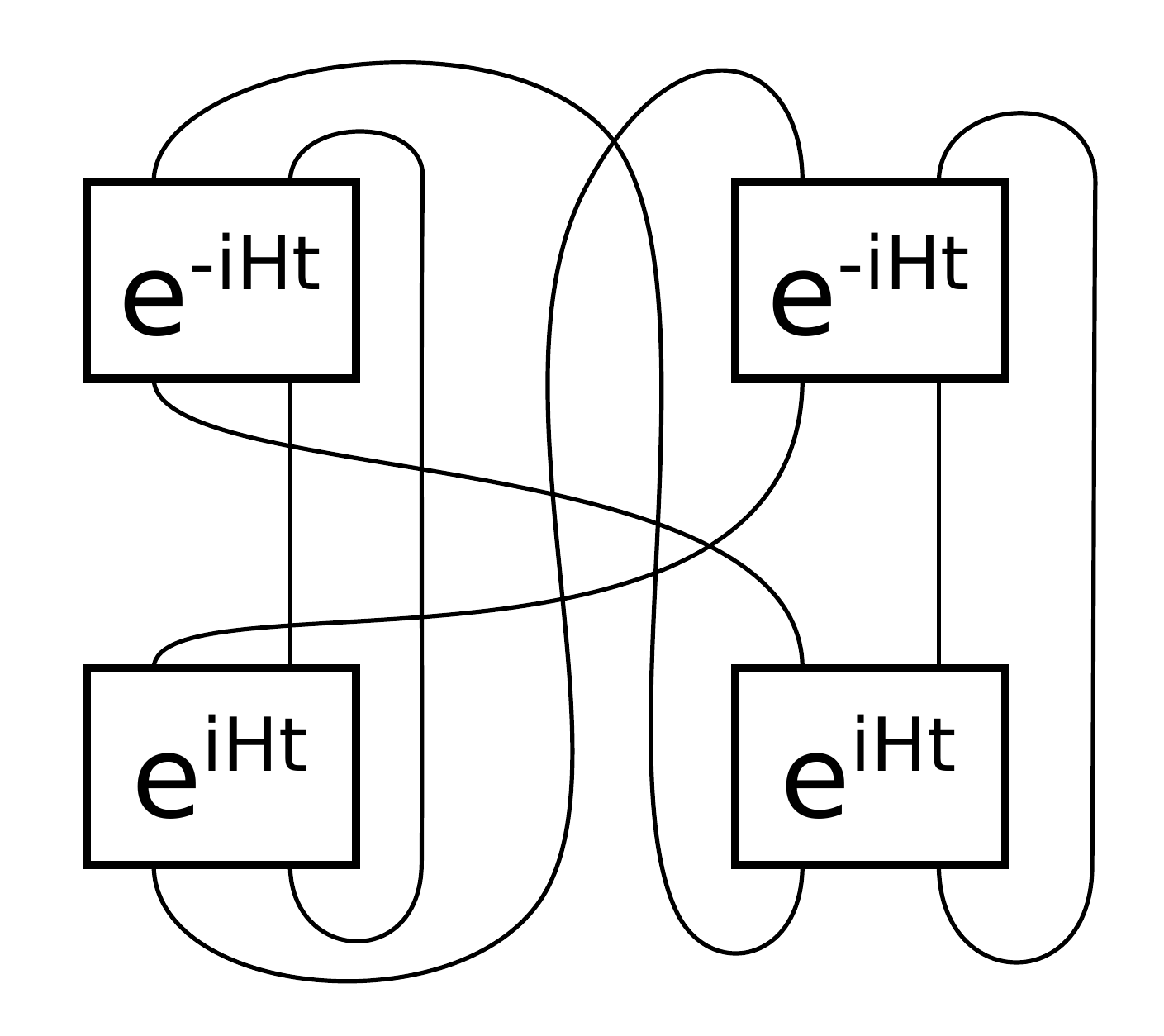}
         \caption{}
         \label{fig:match}
     \end{subfigure}
     \hfill
     \begin{subfigure}[t]{0.45\textwidth}
         \includegraphics[width=\textwidth]{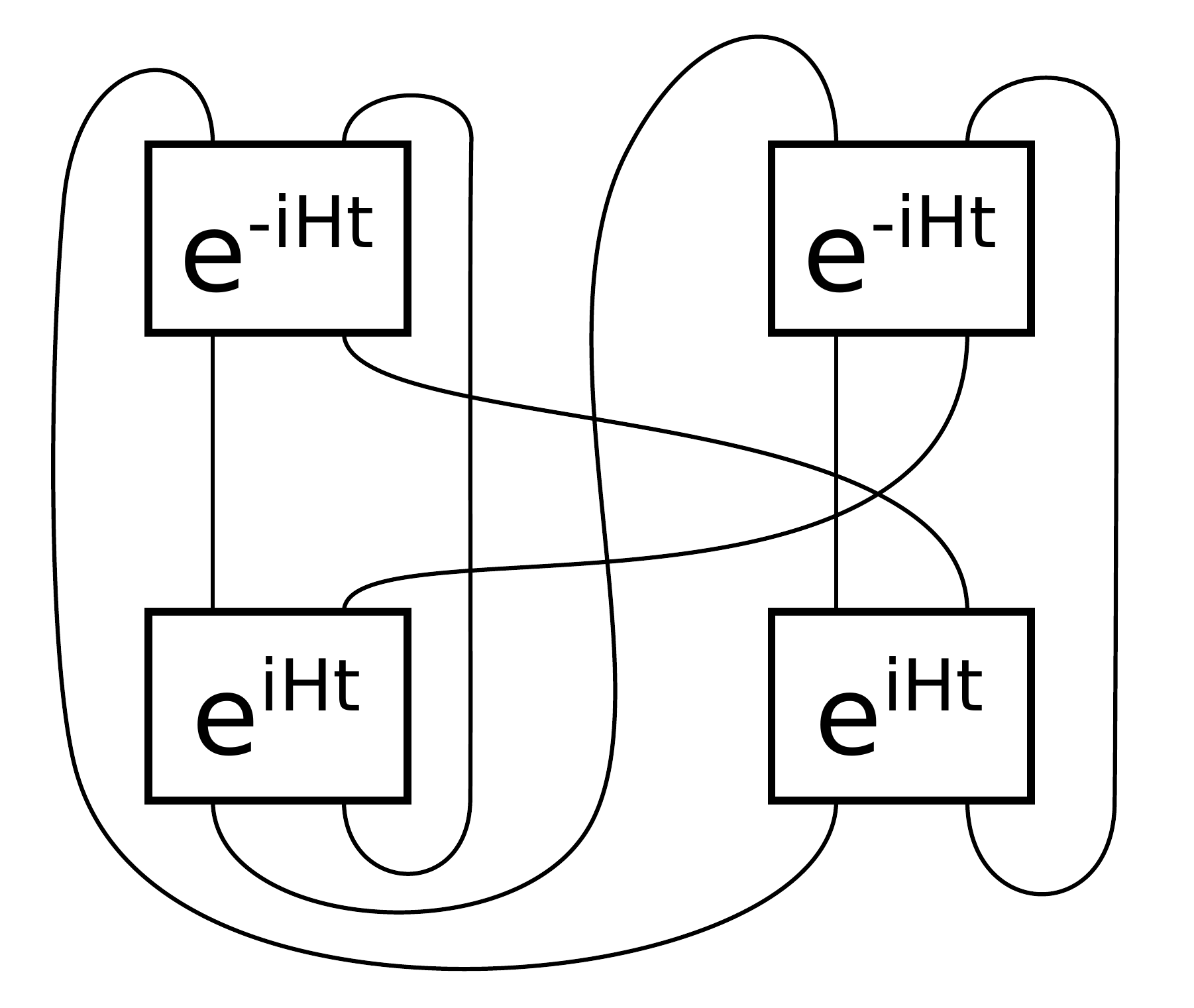}
         \centering
         \subcaption{}
         \label{fig:cross}
     \end{subfigure}
        \caption{Tensor contraction diagrams corresponding to the 2 non-trivial matrix elements found for the case of $k=2$ replicas. Note that the complex conjugate operators were transposed to better fit in the plane, so that the top operators become the inverses of the bottom ones. The convention of having the bottom legs as inputs and top legs as outputs is also reversed for these blocks.}
        \label{fig:contractions}
\end{figure*}

\begin{equation}
    \mathcal{T}^{\star}_{\kappa\epsilon,\sigma\tau} = \bra{\kappa}\otimes\bra{\epsilon} \mathcal{\hat{T}}\ket{\sigma}\otimes \ket{\tau},
\end{equation}
where $\kappa,\,\epsilon,\,\sigma,\,\tau \in S_k$ and $\mathcal{\hat{T}}$ is the diagram corresponding to the unit cell. It is now important to notice that, due to the direct contractions on the legs of the unit cell, all random unitary matrices are contracted with a corresponding conjugate and vanish, so that the matrix element indicated above will take the same value independently of which unitary gates are picked from the Haar ensemble. This means we can ignore Haar averaging in the computation and only focus on the contractions of the core Hamiltonian evolution operators. We note that the only interesting diagrams are those that carry some $t$ dependence, so we need not compute diagrams where the blocks cancel each other. This means we should restrict ourselves to the case of $\kappa \neq \epsilon$ and $\sigma \neq \tau$. For two replicas this condition only leaves 2 nontrivial diagrams corresponding to $\bra{\textbf{S}\textbf{I}}\mathcal{\hat{T}}\ket{\textbf{S}\textbf{I}}$ and $\bra{\textbf{S}\textbf{I}}\mathcal{\hat{T}}\ket{\textbf{I}\textbf{S}}$. These are illustrated as tensor contraction diagrams in Fig. \ref{fig:match} and Fig. \ref{fig:cross} respectively. Since we will eventually take $t$ to be small we can Taylor expand the exponential and compute the diagram perturbatively. It can be easily seen that the linear order correction to both diagrams is just 0, as the contribution from the top blocks exactly cancels that of the bottom blocks. The first non-zero correction to the diagrams appears at order $t^2$. The contributions here come from single boxes in the second order of expansion and from pairs of boxes in the first order of expansion. After summing up all the terms we get the following results:

\begin{equation}
    \begin{split}
        \bra{\textbf{S}\textbf{I}}\mathcal{\hat{T}}\ket{\textbf{S}\textbf{I}} &= d^4 -2t^2\Omega(H)+\mathcal{O}(t^4),\\
        \bra{\textbf{S}\textbf{I}}\mathcal{\hat{T}}\ket{\textbf{I}\textbf{S}} &= d^2 + \mathcal{O}(t^4),
    \end{split}
\end{equation}
where the only dependence on H appears through the function $\Omega(H)$ given in Eq. \ref{eq:omega}. At this level, it appears a coincidence that there is no correction at $t^2$ to the second term. In Appendix \ref{app:multiplereplicas}, we show that similar behavior is found in higher replica calculations.

Knowing these matrix elements, we see that it is possible to recover the propagation tensor using the resolution of identity

\begin{equation}
\begin{split}
    \mathcal{\hat{T}} &= \sum_{\kappa,\epsilon,\sigma,\tau \in S_2} \mathcal{T}^{\star}_{\kappa\epsilon,\sigma\tau}\ket{\kappa^\star}\ket{\epsilon^\star}\bra{\sigma^\star}\bra{\tau^\star} \\
    = &\sum_{\kappa,\epsilon,\sigma,\tau,\mu,\nu \in S_2} \mathcal{W}g(\kappa\mu^{-1}) \mathcal{W}g(\epsilon\nu^{-1}) \mathcal{T}^{\star}_{\kappa\epsilon,\sigma\tau}\ket{\mu}\ket{\nu}\bra{\sigma^\star}\bra{\tau^\star} \\
    = &\sum_{\sigma,\tau,\mu,\nu \in S_2} \mathcal{T}_{\mu\nu,\sigma\tau} \ket{\mu}\ket{\nu}\bra{\sigma^\star}\bra{\tau^\star},
    \end{split}
\end{equation}
where the matrix $\mathcal{T}_{\mu\nu,\sigma\tau}$ defined in the last line is the matrix we are looking for, which tells us how vectors in $V(S_2)\otimes V(S_2)$ evolve under the action of a unit cell. If we let $\mathcal{T}$ and $\mathcal{T}^{\star}$ denote the 4 by 4 matrices with entries defined above, and let $\mathcal{W}g$ be the 2 by 2 matrix with elements $\mathcal{W}g_{\sigma\tau} = \mathcal{W}g(\sigma\tau^{-1})$, then we can compute the transfer matrix knowing the results of all contraction diagrams from the equation

\begin{equation}
\label{eq:Tmatrix}
    \mathcal{T} = \mathcal{W}g\otimes \mathcal{W}g\cdot \mathcal{T}^{\star}.
\end{equation}

Since the unit cell is just the identity for $t=0$, the matrix $\mathcal{T}$ is simply the identity at zeroth order when expanding in powers of $t$. The correction to $\mathcal{T}^\star$ at order $t^2$ was computed above and can be written as

\begin{equation}
\begin{split}
    \Delta^{(2)} \mathcal{T}^\star &= -2t^2\Omega(H)
    \begin{bmatrix}
    0 & 0 & 0 & 0 \\
    0 & 1 & 0 & 0 \\
    0 & 0 & 1 & 0 \\
    0 & 0 & 0 & 0 \\
    \end{bmatrix} \\
    &= -2t^2\Omega(H)\cdot \frac{1}{2}\left(1-\sigma_z\otimes\sigma_z\right).
\end{split}
\end{equation}

If we plug this correction into Eq. \ref{eq:Tmatrix} we get the result presented in Eq. \ref{eq:transfunitcell} of the main text.

\section{Effective dynamics for multiple replicas}
\label{app:multiplereplicas}

Here we give an overview of the method used to compute the transfer matrix using a number of replicas $k>2$. This generalizes the calculation discussed in App. \ref{app:diagrammatic} and it can be reduced to a simple exercise in combinatorics. Suppose we have $\mathcal{\hat{T}}$ the unit cell operator and denote $\bra{\kappa\epsilon}\mathcal{\hat{T}}\ket{\sigma\tau}$ the result of contracting its legs according to permutation operators $\kappa$, $\epsilon$, $\sigma$, $\tau \in S_k$. This will lead to diagrams similar to those seen in Fig. \ref{fig:contractions}, but with $k$ exponential operators of each type. To compute this, we must once again look at the Taylor expansions of the operators. Contributions at order $t^2$ come from the second order term in the expansion of each box and from all pairings of operators expanded to first order. All of these will reduce to one of three possible diagrams, with coefficients of $d$ raised to some power depending on the number of loops that appear. These three are $\tr(H^2)$, $\tr(\tr_1(H)^2)$ and $\tr(H)^2$, the same terms that appear in the definition of $\Omega(H)$.

To find the result of the calculation, consider the following auxiliary construction. Let $i = 1,\ldots,k$ and $\overline{i} = \overline{1},\ldots,\overline{k}$ label a series of boxes and let $\sigma,\,\kappa \in S_k$. Denote the first set of boxes by $\mathcal{B}$ and the second by $\overline{\mathcal{B}}$. Now imagine that we connect the bottom of the $i$ boxes to the top of the $\overline{i}$ boxes according to the map $i\to\overline{\sigma(i)}$, and similarly the top of the $i$ boxes to the bottom of the $\overline{i}$ boxes according to $i\to\overline{\kappa(i)}$. The result is that the boxes are now tied together in chains of varying lengths, each chain containing an equal number of $i$ and $\overline{i}$ boxes. If we call $l$ the number of chains, then we let $\mu = 1,\ldots,l$ index the chains and form sets $\mathcal{S}_\mu$ as

\begin{equation}
    \mathcal{S}_\mu = \{b \in \mathcal{B}\cup\overline{\mathcal{B}}\mid\ b\text{ is in chain }\mu\}.
\end{equation}

These sets exactly partition the full set of boxes with no overlaps

\begin{equation}
    \bigcup_\mu \mathcal{S}_\mu = \mathcal{B}\cup \overline{\mathcal{B}}.
\end{equation}

We now return to our initial problem and consider the sets $\mathcal{S}^{L}_\mu$ and $\mathcal{S}^{R}_\nu$ formed by the left $(\sigma,\,\kappa)$ and right $(\tau,\,\epsilon)$ pairs of permutations. The indices are allowed to go from 1 to the number of chains formed by the respective pair. Denote by $\mathcal{M}_{\mu\nu}$ the number of elements from $\mathcal{B}$ that are found in both $\mathcal{S}^{L}_\mu$ and $\mathcal{S}^{R}_\nu$, and by $\overline{\mathcal{M}}_{\mu\nu}$ the number of elements from $\overline{\mathcal{B}}$ found in the same 2 partitions. Remarkably, counting the number of diagrams of the three types mentioned above that show up in the transfer matrix reveals that the correction to second order in $t$ is given by the simple expression

\begin{equation}
    \Delta^{(2)} \mathcal{T}^\star_{\kappa\epsilon,\sigma\tau} = -\frac{1}{2d^4}t^2\Omega(H) \braket{\kappa | \sigma}\braket{\epsilon | \tau}\norm{\mathcal{M}-\overline{\mathcal{M}}}^2_2.
\end{equation}

This result shows that even for a higher number of replicas, the characteristics of the evolution are still independent of the microscopic Hamiltonian $H$, which only sets the overall timescale through the same function $\Omega(H)$ as in the $k=2$ case. The interactions produced by this transfer matrix may have an interesting interpretation in terms of domain walls, but a discussion of this is beyond the scope of the current work.

\section{Evolution of the $\mathbf{X}$ state}
\label{app:Xstate}

Following a similar calculation of the matrix elements in the transfer matrix as shown in App. \ref{app:diagrammatic} we can also derive the unitary evolution of the $\mathbf{X}$ states that appear after a measurement. If we introduce a projection operator onto this state $P_\mathbf{X}$, such that $P_\mathbf{X}\ket{\mathbf{X}} = \ket{\mathbf{X}}$ and $P_\mathbf{X} V(S_2) = 0$ and denote by $Q_\mathbf{X}$ its complement so that $P_\mathbf{X}+Q_\mathbf{X} = I$, the total effective Hamiltonian is given by
\begin{equation}
\begin{split}
    \mathcal{H}^{\mathcal{M}}_{ij} = &\mathcal{H}_{ij}Q_{\mathbf{X}}^{(i)}Q_{\mathbf{X}}^{(j)}+\mathcal{H}_{\mathbf{X}}^{(j)} P_{\mathbf{X}}^{(i)}Q_{\mathbf{X}}^{(j)} \\
    &+ \mathcal{H}_{\mathbf{X}}^{(i)} Q_{\mathbf{X}}^{(i)}P_{\mathbf{X}}^{(j)} + E_{\mathbf{X}\mathbf{X}} P_{\mathbf{X}}^{(i)}P_{\mathbf{X}}^{(j)},
\end{split}
\end{equation}
where $\mathcal{H}_{ij}$ is the restriction to $V(S_2)$ of the effective Hamiltonian in Eq. \ref{eq:hamiltonianij} of the main text, the operator $\mathcal{H}_{\mathbf{X}}^{(i)}$ is given by
\begin{equation}
\label{eq:hamiltX}
    \mathcal{H}_{\mathbf{X}}^{(i)} = \Gamma \frac{(d+1)(d^2-1)}{d^3}+\gamma (1-\frac{d+1}{d^3}) + \frac{\gamma}{d^2}\sigma_x,
\end{equation}
and the $\mathbf{X}\mathbf{X}$ interaction energy is
\begin{equation}
\label{eq:energyX}
    E_{\mathbf{X}\mathbf{X}} = \frac{\gamma}{d^2}\left(1-\frac{2}{d(d^2-1)}\right)+2\frac{\Gamma}{d^2}\left(1+\frac{1+d^2}{d(d^2-1)}\right).
\end{equation}

We see that the only effect of $\Gamma$ is to raise the energy of configurations that include $\mathbf{X}$ states. It should then be intuitively clear that taking a very large $\Gamma$ will result in the dynamics being projected into the $\mathbf{X}$-free subspace of the Hilbert space. The rest of the appendix constitutes a formal proof of this fact.

Denote by $\mathcal{P}$ the projector onto the $\mathbf{X}$-free subspace of the Hilbert space and $\mathcal{Q} = I-\mathcal{P}$ its complement. To simplify notation let $\mathcal{H}$ denote the full effective Hamiltonian over the course of this proof. This includes interactions of the type $\mathcal{H}^\mathcal{M}_{ij}$ above between all nearest-neighbors in the chain and selective measurements $\mathcal{M}$ at some frequency $f$ on all sites (defined in Eq. \ref{eq:measurementop}). Then we can separate $\mathcal{H}$ as
\begin{equation}
\begin{split}
    \mathcal{H}&= (\mathcal{P}\mathcal{H}\mathcal{P} + \mathcal{Q}\mathcal{H}\mathcal{Q}) + (\mathcal{P}\mathcal{H}\mathcal{Q}+\mathcal{Q}\mathcal{H}\mathcal{P}) \\
    &= \mathcal{H}_0+\Delta \mathcal{H},
\end{split}
\end{equation}
where the diagonal parts become $\mathcal{H}_0$ and the coupling between sectors is $\Delta \mathcal{H}$. If we let $\mathcal{U} = \exp(-t\mathcal{H})$ be the imaginary time propagator then its evolution equation is
\begin{equation}
    \frac{d}{dt}\mathcal{U} = -\mathcal{H}\mathcal{U}.
\end{equation}

We move into the interaction picture of $\mathcal{H}_0$ by letting $\mathcal{U}(t) = \exp(-t\mathcal{H}_0)\mathcal{U}_I(t)$. The evolution equation of $\mathcal{U}_I(t)$ is
\begin{equation}
    \frac{d}{dt}\mathcal{U}_I(t) = -\Delta \mathcal{H}_I(t) \mathcal{U}_I(t),
\end{equation}
where
\begin{equation}
    \Delta \mathcal{H}_I(t) = e^{t\mathcal{H}_0} \Delta \mathcal{H} e^{-t\mathcal{H}_0}.
\end{equation}

The boundary conditions on all quantities of interest lie in the $\mathbf{X}$-free subspace of the Hilbert space and configurations that include $\mathbf{X}$ states have 0 overlap with this sector. Therefore, we are only interested in the reduced propagator $\mathcal{P}\mathcal{U}_I(t)\mathcal{P}$. This evolves according to
\begin{equation}
    \frac{d}{dt} \mathcal{P}\mathcal{U}_I(t)\mathcal{P} = -\mathcal{P}e^{t\mathcal{H}_0}P\Delta \mathcal{H}\mathcal{Q}e^{-t\mathcal{H}_0}\mathcal{Q}\mathcal{U}_I(t)\mathcal{P}.
\end{equation}

We see that the evolution equation requires knowledge about $\mathcal{Q}\mathcal{U}_I(t)\mathcal{P}$. This evolves according to
\begin{equation}
    \frac{d}{dt} \mathcal{Q}\mathcal{U}_I(t)\mathcal{P} = -\mathcal{Q}e^{t\mathcal{H}_0}Q\Delta \mathcal{H}\mathcal{P}e^{-t\mathcal{H}_0}\mathcal{P}\mathcal{U}_I(t)\mathcal{P}.
\end{equation}

We integrate the two equations from $0$ to $t$ and substitute the form of $\mathcal{Q}\mathcal{U}_I(t)\mathcal{P}$ into the $\mathcal{P}\mathcal{U}_I(t)\mathcal{P}$ equation to arrive at the integral form
\begin{equation}
\label{eq:propagator}
\begin{split}
    \mathcal{P}\mathcal{U}_I(t)\mathcal{P} &= \mathcal{P}-\int_0^t\int_0^{t'}dt' dt'' \mathcal{P}e^{t'\mathcal{H}_0}\mathcal{P}\Delta \mathcal{H}\\
    &\times \mathcal{Q}e^{-(t'-t'')\mathcal{H}_0} \mathcal{Q}\Delta \mathcal{H} \mathcal{P} \mathcal{U}_I(t'')\mathcal{P} \\
    &=\mathcal{P} - \delta \mathcal{U}(t)
\end{split}
\end{equation}

Let us now look at the operator norm of the correction term on the RHS. Since this is both sub-additive and sub-multiplicative we have
\begin{equation}
\begin{split}
    \norm{\delta \mathcal{U}(t)} &\leq \int_0^t\int_0^{t'}dt' dt'' \norm{\mathcal{P}e^{t'\mathcal{H}_0}\mathcal{P}\Delta \mathcal{H} \mathcal{Q}} \\
    &\times \norm{\mathcal{Q}e^{-(t'-t'')\mathcal{H}_0} \mathcal{Q}} \times\norm{\mathcal{Q}\Delta \mathcal{H} \mathcal{P}}\times\norm{ \mathcal{P}\mathcal{U}_I(t'')\mathcal{P}}. \\
\end{split}
\end{equation}

Our aim is now to find a bound for each norm in the expression. The connection between the different sectors $\Delta \mathcal{H}$ is bound and independent of $\Gamma$, as it appears only due to the measurements and is not a result of the random unitaries. The $\mathcal{P}\mathcal{H}_0\mathcal{P}$ drives evolution in the $\mathbf{X}$-free subspace, which was also shown to be independent of $\Gamma$. Therefore we define

\begin{equation}
    m(t) = \sup_{t'\in (0,t)}\left(\norm{\mathcal{Q}\Delta \mathcal{H} \mathcal{P}} \norm{\mathcal{P}e^{t'\mathcal{H}_0}\mathcal{P}\Delta \mathcal{H} \mathcal{Q}}\right),
\end{equation}
a real and continuous function. Note that $m(t)$ is strictly positive and non-decreasing by construction. Furthermore, it is independent of $\Gamma$ by the previous arguments. For the $\mathcal{Q}\mathcal{H}_0\mathcal{Q}$ part of the Hamiltonian, the energy of all configurations is raised by some constant proportional to $\Gamma$, as it is clear from Eq. \ref{eq:hamiltX} and Eq. \ref{eq:energyX}. Therefore there must exist positive constants $a,b$ such that the following inequality holds
\begin{equation}
    \norm{\mathcal{Q}e^{-(t'-t'')\mathcal{H}_0} \mathcal{Q}} \leq e^{-(t'-t'')(a\Gamma-b)},
\end{equation}
for all $t' > t''$. The energy $E(\Gamma) = a\Gamma-b $ can be interpreted as a lower bound on the ground state energy of the restricted Hamiltonian $\mathcal{Q}\mathcal{H}_0 \mathcal{Q}$. We are only interested in the large $\Gamma$ limit, so we will assume $E(\Gamma)>0$. For the final part we have
\begin{equation}
    \norm{ \mathcal{P}\mathcal{U}_I(t'')\mathcal{P}} \leq \norm{P}+\norm{\delta\mathcal{U}(t'')} = 1+\norm{\delta\mathcal{U}(t'')},
\end{equation}
since the norm of a projector is always $1$. If we introduce the notation $\psi (t) = \norm{\delta\mathcal{U}(t)}/m(t)$ and combine all of the previous inequalities, we have
\begin{equation}
    \psi(t) \leq \int_0^t\int_0^{t'}dt' dt'' e^{-(t'-t'')E(\Gamma)}(1+m(t'')\psi(t'')).
\end{equation}

To proceed, we invert the order of the integrals, so that
\begin{equation}
    \psi(t) \leq \int_0^t dt''\int_{t''}^{t}dt' e^{-(t'-t'')E(\Gamma)}(1+m(t'')\psi(t'')).
\end{equation}

We can now perform the integration over $t'$. A quick calculation shows that

\begin{equation}
    \int_{t''}^{t}dt' e^{-(t'-t'')E(\Gamma)} \leq \frac{1}{E(\Gamma)}.
\end{equation}

Adding this to the above and renaming the dummy variable $t'' \to s$ we arrive at
\begin{equation}
    \psi(t) \leq \frac{t}{E(\Gamma)}+\int_0^t ds \frac{m(s)}{E(\Gamma)}\psi(s).
\end{equation}

We now use the integral form of Grönwall's inequality to get the bound on $\psi(t)$
\begin{equation}
    \psi(t) \leq \frac{t}{E(\Gamma)}\exp(\int_0^t \frac{m(s)}{E(\Gamma)}ds),
\end{equation}
and from the definition of $\psi(t)$ we get
\begin{equation}
    \norm{\delta \mathcal{U}(t)} \leq \frac{tm(t)}{E(\Gamma)}\exp(\int_0^t \frac{m(s)}{E(\Gamma)}ds).
\end{equation}

We can now take the limit of $\Gamma \to \infty$ and use the positivity of the norm to get

\begin{equation}
    \lim_{\Gamma \to \infty} \norm{\delta \mathcal{U}(t)} = 0,
\end{equation}
which implies that the operator itself must become null in this limit. From Eq. \ref{eq:propagator}, we then have that the interaction picture propagator acts as identity on the $\mathbf{X}$-free subspace in this limit. Returning to the original frame, we see that the propagator must then be
\begin{equation}
    \mathcal{U}(t) = \exp (-t\mathcal{H}_0)\mathcal{P} = \exp (-t \mathcal{P} \mathcal{H}\mathcal{P}),
\end{equation}
and the dynamics are driven by the restriction of the Hamiltonian $\mathcal{P}\mathcal{H}\mathcal{P}$, as claimed in the main text.

\section{Connection between the microscopic and the effective Hamiltonian}
\label{app:micromacro}

In Sec. \ref{sec:unitmodel} we saw that $\gamma$ sets the timescale for information transfer across the network. To deduce the meaning of $\Gamma$, defined in Eq.~\eqref{eq:Gamma}, we return to the microscopic Hamiltonian $H$ and expand it in a convenient qudit basis called the generalized Pauli group \cite{Gottesman1999}. This is generated by operators $X_d$, $Z_d$ defined through their action on the basis states
\begin{align}
    X_d \ket{j} &= \ket{j\oplus1}, \\
    Z_d \ket{j} &= \omega^j \ket{j},
\end{align}
where $\omega$ is a primitive $d$'th root of unity and addition is understood to be modulo $d$. These obey the braiding equation
\begin{equation}
\label{eq:braiding}
    \left(X_{d}^{a} Z_{d}^{b}\right)\left(X_{d}^{s} Z_{d}^{t}\right)=\omega^{b s-a t}\left(X_{d}^{s} Z_{d}^{t}\right)\left(X_{d}^{a} Z_{d}^{b}\right).
\end{equation}

We can then expand the Hamiltonian on this basis as
\begin{equation}
    H = \sum_{a,b,s,t = 0}^{d-1} h_{ab;st} X_d^a Z_d^b \otimes X_d^s Z_d^t.
\end{equation}

It is convenient to separate the Hamiltonian into an entangling and a non-entangling sector. We define the non-entangling sector of the basis to be spanned by those operators which act trivially as identity on at least one of the qudits, and the entangling sector is everything else. We can write this split as
\begin{equation}
    H = H_{loc}+H_{int},
\end{equation}
and equivalently for the matrices
\begin{equation}
    h = h_{loc}+h_{int}.
\end{equation}

In a microscopic physical picture the $H_{int}$ stores information about the nearest neighbor interactions in the chain, while $H_{loc}$ generates local dynamics such as interactions with magnetic fields.

For simplicity, we can assume $\tr(H) = 0$ such that $h_{00;00}=0$. Additionally, since we assumed $H$ is symmetric under swapping the two qudits, we get
\begin{equation}
\label{eq:hermiticity}
    h_{ab;st} = h_{st;ab},
\end{equation}
and the hermiticity condition $H = H^{\dagger}$ of the Hamiltonian implies
\begin{equation}
    h_{-a-b;-s-t} = h^*_{ab;st}\omega^{ab+st},
\end{equation}
where the inverse $-a$ is with respect to modulo $d$ addition $a\oplus(-a) = 0$. We can start by calculating $\Gamma$ for this Hamiltonian. If we notice that all elements in our basis other than $I$ are traceless, we have
\begin{equation}
\begin{split}
    \tr_1(H) &= \sum_{a,b,s,t = 0}^{d-1} h_{ab;st} X_d^s Z_d^t \tr(X_d^a Z_d^b) \\
    &= d \sum_{s,t = 0}^{d-1} h_{00;st} X_d^s Z_d^t.
\end{split}
\end{equation}

If we now square this result and apply the braiding equation \ref{eq:braiding} we get
\begin{equation}
\begin{split}
    \tr_1(H)^2 &= d^2 \sum_{a,b,s,t = 0}^{d-1} h_{00;ab} h_{00;st} X_d^a Z_d^b X_d^s Z_d^t \\
    &= d^2 \sum_{a,b,s,t = 0}^{d-1} h_{00;ab} h_{00;st} \omega^{at} X_d^{a\oplus s} Z_d^{b\oplus t}.
\end{split}
\end{equation}

Finally, we take the trace of this and use Eq. \ref{eq:hermiticity} to get
\begin{equation}
\begin{split}
    \Gamma &= \frac{2d}{(d^2-1)^2}\tr(\tr_1(H)^2)\\ &= \frac{2d^3}{(d^2-1)^2} \sum_{a,b,s,t = 0}^{d-1} h_{00;ab} h_{00;st} \omega^{at} \tr(X_d^{a\oplus s} Z_d^{b\oplus t}) \\
    &= \frac{2d^4}{(d^2-1)^2} \sum_{a,b,s,t = 0}^{d-1} h_{00;ab} h_{00;st} \omega^{at} \delta_{a\oplus s,0}\delta_{b \oplus t,0} \\
    &= \frac{2d^4}{(d^2-1)^2} \sum_{a,b = 0}^{d-1} h_{00;ab} h_{00;-a-b} \omega^{-ab} \\
    &= \frac{2d^4}{(d^2-1)^2} \sum_{a,b = 0}^{d-1} \abs{h_{00;ab}}^2.
\end{split}
\end{equation}

We can write this more compactly in terms of the Frobenius norm of the local part of the Hamiltonian
\begin{equation}
    \Gamma = \frac{d^4}{(d^2-1)^2}\norm{h_{loc}}_{\mathrm{F}}^2,
\end{equation}
with a coefficient that goes to 1 in the $d \to \infty$ limit. It is clear that this part of the Hamiltonian is only relevant in randomizing the state of individual qudits after measurement and does not play any role in the transfer of information across the chain.

For the information transfer rate $\gamma$, we can compute
\begin{equation}
\begin{split}
    \tr(H^2) &= d^2\sum_{a,b,s,t = 0}^{d-1} h_{ab;st} h_{-a-b;-s-t} \omega^{-ab-st} \\
    &= d^2\sum_{a,b,s,t = 0}^{d-1} \abs{h_{ab;st}}^2,
\end{split}
\end{equation}
so from Eq. \ref{eq:omega} we have
\begin{equation}
\begin{split}
    \frac{\gamma}{2} &= \frac{d^2}{(d^2-1)^2}\tr(H^2)-\frac{2d}{(d^2-1)^2}\tr(\tr_1(H)^2) \\
    &= \frac{d^4}{(d^2-1)}\left(\sum_{a,b,s,t = 0}^{d-1} \abs{h_{ab;st}}^2 - 2\sum_{a,b = 0}^{d-1} \abs{h_{00;ab}}^2\right) \\
    &= \frac{d^4}{(d^2-1)^2} \norm{h_{int}}^2_\mathrm{F},
\end{split} 
\end{equation}
showing that only the interaction part of the Hamiltonian leads to transfer of information, as expected.

\section{Evaluation of $\Theta(t)$ for periodic boundary conditions }
\label{app:pbc}

With periodic boundary conditions, the non-interacting fermionic Hamiltonian introduced in Eq.~\eqref{eq:HamFermionic} can be block diagonalized in terms of states with definite quasimomentum. Working in units where $\gamma = 1$, a standard computation gives
\begin{equation}
    H_p = -\frac{1}{2}\sum_{k\in \mathcal{K}_p} 2(\cos k-g)a_k^\dagger a_k + ( e^{\i k} a^\dagger_k a^{\dagger}_{-k} + \text{H.c.}) + g.
\end{equation}
where $a_k = \sum_{j}e^{-\iu k j} a_j$ are annihilation operators for fermions with wavevector $k$.
The set of $k$-space points $\mathcal{K}_p$ in the above sum depends on the fermion parity sector $p$ \cite{Mbeng2020}. Assuming an even number of sites $N$, the odd parity sector $p = 1$ has periodic boundaries (PBCs) $e^{i k N} = +1$ while the even parity sector $p = 0$ has antiperiodic boundaries (ABCs) $e^{i k N} = -1$. Since the anomalous terms pair $+k$ and $-k$ modes, it is helpful to single out the positive $k$-modes
\begin{align}
    \mathcal{K}_{p=1}^+ &= \{k = 2\pi n/L, n = 1, \ldots, (L/2)-1\} \\
    \mathcal{K}_{p=0}^+ &= \{k = (2n-1)\pi/L, n = 1, \ldots, (L/2)\}
\end{align}
Note that the $p = 1$ case has one fewer $k$ value, which is resolved by accounting for modes $k = 0$ and $k = \pi$, for which the anomalous terms vanish
\begin{align}
    H_{k=0,\pi} = -(1-g)a_0^\dagger a_0 + (1+g)a^\dagger_\pi a_\pi - g
\end{align}
Overall, this allows us to write the Hamiltonian in matrix form
\begin{align}
    H_p = -\sum_{k\in \mathcal{K}_p}
    \begin{pmatrix}
        a_k^\dagger & a_{-k}
    \end{pmatrix}
    \mathcal{H}_k
    \begin{pmatrix}
        a_k \\ a^\dagger_{-k}
    \end{pmatrix}
\end{align}
We can write $\mathcal{H}_k = \vec{h}(k)\cdot \vec{\sigma}$, where $\vec{\sigma}$ is a 3-vector of Pauli matrices, and we have
\begin{align}
    \vec{h}(k) = \begin{bmatrix}
        0 \\ -\sin k \\ \cos k-g
    \end{bmatrix}
\end{align}
When this $k$-space Hamiltonian $\mathcal{H}_k$ is diagonalized, we obtain energies
\begin{align}
    \epsilon_k = \pm \sqrt{1+g^2 - 2g\cos k},
\end{align}
consistent with the dispersion relation \eqref{eq:Dispersion} quoted in the main text. As a reminder, the object we wish to calculate is 
\begin{align}
    \Theta = \frac{\Tr[\ket{+}\bra{+} e^{-t H_{p=0}}]}{\Tr[\ket{-}\bra{-} e^{-t H_{p=1}}]}
    \label{eq:ThetaHK}
\end{align}
Now, we note that the state $\ket{+}$ ($\ket{-}$) is the ground state of the Hamiltonian $\mathcal{H}_{\rm eff}$ in the even (odd) parity sector, for the specific case $g = 0$. Both objects in each of the traces in Eq.~\eqref{eq:ThetaHK} are $k$-diagonal, and so both the numerator and denominator of the above become a product over $k$. Each factor of the product is given by an overlap between two states of a two-level system spanned by $\ket{\text{VAC}}$ and $a^\dagger_k a^\dagger_{-k}\ket{\text{VAC}}$, namely the overlap between the ground state with $g = 0$ and the thermal state with temperature $t$ for nonzero $g$. We should be careful to include the modes $k = 0, \pi$, that are present in the denominator, separately. 

Denoting $\rho_{+,k}$ as the (pure) $2\times 2$ density matrix for mode $k$ in the state $\ket{+}$, we have $\rho_{+,k} = (I_2 + \vec{n}_{+,k}\cdot \vec{\sigma})/2$, where
\begin{align}
    \vec{n}_{+,k} = \begin{bmatrix}
        0 \\ -\sin k \\ \cos k
    \end{bmatrix}
\end{align}
Meanwhile, using the diagonal representation of $\mathcal{H}_k$ we get
\begin{align}
    e^{-t \mathcal{H}_k} &= e^{+\epsilon_k t}\frac{1}{2}(I_2+\vec{n}_{H,k}\cdot \vec{\sigma}) + e^{-\epsilon_k t}\frac{1}{2}(I_2-\vec{n}_{H,k}\cdot \vec{\sigma}) \nonumber\\ &= 2\cosh(\epsilon_k t)\left(\frac{I_2 + \tanh(\epsilon_k t)\vec{n}_{H, k}\cdot \vec{\sigma}}{2}\right)
\end{align}
where
\begin{align}
    \vec{n}_{H, k} = \frac{1}{\epsilon_k}\vec{h}(k)
\end{align}
is the 3-vector that specifies the ground state of the 2-level Hamiltonian $\mathcal{H}_k$.  Now, since
\begin{align}
    \Tr\left[\left(\frac{I_2 + \vec{n}_1\cdot \vec{\sigma}}{2}\right)\cdot \left(\frac{I_2 + \vec{n}_2\cdot \vec{\sigma}}{2}\right)\right] = \frac{1 + \vec{n}_1 \cdot \vec{n}_2}{2}
\end{align}
we can compute the dot product $\vec{n}_{+,k} \cdot \vec{n}_{H,k} = (\sin^2k + \cos k(\cos k-g))/\epsilon_k = (1-g\cos k)/\epsilon_k$, and the relevant factor for each $k$ becomes
\begin{align}
    \theta(k,t) = \cosh(\epsilon_k t)\left[1 + \tanh( \epsilon_k t)\frac{1-g \cos k}{\epsilon_k}\right]
\end{align}
Finally, the relevant factor for the $k = 0, \pi$ modes comes from recognizing that in the state $\ket{-}$, we have $a^\dagger_0 a_0 = 1$ and $a^\dagger_\pi a_\pi = 0$. Because $H_{p=0}$ conserves these occupations, we get a factor of $e^{t}$ in the numerator. Overall, we obtain the expression
\begin{align}
    \Theta = e^{t} \frac{\prod_{k \in \mathcal{K}_{p=0}^+}\theta(k,t)}{\prod_{k \in \mathcal{K}_{p=1}^+}\theta(k,t)}
\end{align}
which was quoted in the main text, Eq.~\eqref{eq:ThetaProductPBC}.

Now, to evaluate the products in the above, we define the function
\begin{equation}
    f(z) = \frac{\log \theta(z, t)}{\sin Nz}
\end{equation}
where $z$ is now a complex variable which corresponds to the wavevctor $k$ on the real axis. If one considers the integral of $f(z)$ around a thin contour $\Gamma$ encircling the real interval $[-\pi,\pi]$, cutting the line at the edges at a straight angle, then using the residue theorem we can show that
\begin{equation}
    \log \Theta = \frac{N}{4\pi i}\oint_\Gamma dz f(z).
\end{equation}
By deforming the integration contour, the above can be re-expressed as
\begin{equation}
    \oint_\Gamma dz f(z) = \int_{-\infty}^{\infty}idx f(ix-\delta) - \int_{-\infty}^{\infty}idx f(ix+\delta),
\end{equation}
which integrates on both sides of the branch cut chosen along the imaginary line. We do not include it explicitly but we keep in mind that the contour has a small gap such as to not cross the real line. Consider the rotated $\theta$ function $\theta_r(x) = -i\theta(ix)$ (we leave the $t$-dependence implicit), and equivalently for the energy function $\epsilon_r$. We have
\begin{equation}
\begin{split}
    \theta_r(x) &= \cos t\epsilon_r(x) - \sin t\epsilon_r \frac{g\cosh x-1}{\epsilon_r(x)} \\
    &=\frac{\cos(t\epsilon_r(x)+\phi(x))}{\cos \phi(x)},
\end{split}
\end{equation}
where we introduced $\tan \phi(x) = (g\cosh(x)-1)/\epsilon_r(x)$. By symmetry we can focus on the integral above the real line and have
\begin{equation}
\begin{split}
    \frac{1}{2}\oint dz f(z) &= i\int_K^\infty dx \left[f(ix-\delta)-f(ix+\delta)\right] \\
    &= \int_K^\infty \frac{dx}{\sinh(Nx)}\log\frac{\theta_r(x+i\delta)}{\theta_r(x-i\delta)},
\end{split}
\end{equation}
where $K = -\log g$ is the point where $\epsilon_r(x) = 0$. Consider the Weierstrass factorization of the cosine
\begin{equation}
\cos z=\prod_{q \in \mathbb{Z}, q \text { odd }}\left(1-\frac{2 z}{q\pi}\right) e^{2 z / q\pi},
\end{equation}
and apply it to $\theta_r$. The integral will exactly cancel for all terms that are analytic in the top half plane, so we only need to keep the factors that have a 0 above $K$ on the imaginary line. The $0$ at $K$ is exactly cancelled by the denominator in the expression of $\theta_r$. Then the integral is
\begin{equation}
\begin{split}
    &\frac{1}{2}\int dz f(z) = \int_K^{\infty} \frac{dx}{\sinh Nx} \\
    &\times \sum_{q>0, q \text{ odd}}\log \frac{1-\frac{2}{q\pi}(t\epsilon_r(x+i\delta)+\phi(x+i\delta))}{1-\frac{2}{q\pi}(t\epsilon_r(x-i\delta)+\phi(x-i\delta))}.
\end{split}
\end{equation}

Since the function $t\epsilon_r(x)+\phi(x)$ is strictly increasing from $-\pi/2$ to $\infty$ as $x$ goes from 0 to $\infty$ each of the $q$ terms will have exactly one node on the integration line. The value of the logarithm differs by exactly $2\pi i$ starting from the node up to $\infty$. If we denote by $x_q$ the unique solution of the equation $t\epsilon(x)+\phi(x) = q\pi/2$, the value of the integral is
\begin{equation}
\begin{split}
    \frac{1}{2}\int dz f(z) &= -2\pi i\sum_{q>0,q\text{ odd}} \int_{x_q}^{\infty} \frac{dx}{\sinh Nx}  \\
    &=\frac{2\pi i}{N}\sum_q \log \tanh \frac{x_q N}{2}.
\end{split}
\end{equation}

Then we see that
\begin{equation}
    \log \Theta = \sum_{q \text{ odd}} \tanh \frac{Nx_q}{2}
\end{equation}
as claimed in the main text.

\section{Product representation of $\Theta$ for open boundary conditions}
\label{app:prodrep}

To compute the determinants that appear in Eq. \ref{eq:Theta}, we need an explicit representation for the eigenvector matrix $\mathcal{O}$ of the grand dynamical matrix $\mathcal{D}$. Due to the particle-hole symmetry, this can be written in the form
\begin{equation}
    \mathcal{O} = 
    \begin{bmatrix}
    u & v \\
    v & u,
    \end{bmatrix},\;
    \mathcal{O}^{-1} = \mathcal{O}^T.
\end{equation}

If we let $i,\,j = \overline{1,N-1}$ and denote $m = N/2$ (assuming even $N$)
\begin{equation}
    u = \begin{bmatrix}
    0 & u_{0j} \\
    u_{i0} & u_{ij},
    \end{bmatrix},\; 
    v = \begin{bmatrix}
    v_{00} & v_{0j} \\
    v_{i0} & v_{ij}.
    \end{bmatrix}
\end{equation}

Let $y = i-m$ be the coordinate relative to the middle. Then the eigenvectors can be represented as
\begin{equation}
    u_{ij} = \left\{\begin{array}{lr}
        a_j \frac{\cos k_j y}{\cos k_j m}, & \text{for } j \text{ odd}\\
        a_j \frac{\sin k_j y}{\sin k_j m} & \text{for } j \text{ even}
        \end{array} \right.,
\end{equation}
\begin{equation}
    u_{0j} = \left\{\begin{array}{lr}
        2a_j, & \text{for } j \text{ odd}\\
        0, & \text{for } j \text{ even}
        \end{array} \right.,
\end{equation}
\begin{equation}
    v_{ij} = \left\{\begin{array}{lr}
        a_j \frac{\sin k_j y}{\sin k_j m}, & \text{for } j \text{ odd} \\
        a_j \frac{\cos k_j y}{\cos k_j m}, & \text{for } j \text{ even}
        \end{array} \right.,
\end{equation}
\begin{equation}
    v_{0j} = \left\{\begin{array}{lr}
        0, & \text{for } j \text{ odd} \\
        2a_j, & \text{for } j \text{ even}
        \end{array} \right.,
\end{equation}
\begin{equation}
    v_{i0} = a_0 \frac{\cos k_0 y}{\cos k_0 m},\;
    u_{i0} = a_0 \frac{\sin k_0 y}{\sin k_0 m},
\end{equation}
\begin{equation}
    v_{00} = 2a_0,
\end{equation}

The normalization constants $a_j$ are chosen such that for all $j$ (and the additional $j=0$ mode)
\begin{equation}
    4 a_j^2 + \sum_{i} \left(u_{ij}^2+v_{ij}^2\right)=1,
\end{equation}
although their explicit values are not important in our calculation.

The wavevectors $k_j$ are in $[0, \pi)$ (except for $k_0$, which becomes imaginary at the transition) and are the solutions of the equation
\begin{equation}
    \tan k_jN = \frac{g\sin k_j}{1-g\cos k_j}.
\end{equation}

The corresponding eigenvalues are given by the well-known dispersion relation for the TFIM
\begin{equation}
    \lambda^2 = 1+g^2-2g \cos{k}.
\end{equation}

Since the matrices $u$ and $v$ are split into even and odd sectors one can show there must be some linear dependence, such that $\det u = \det v = 0$. This problem disappears when looking at the $i,j \geq 1$ sector, such that the inverses $u^{-1}_{ij}$ and $v^{-1}_{ij}$ exist. We can then proceed to calculate the determinants in the expression for $\Theta$. We saw that at long enough times, we can make the approximation
\begin{equation}
    \Gamma[-t\mathcal{D}] \approx \mathcal{O}\begin{bmatrix}
    0 & 0 & 0 &  0 \\
    0 & -I_{N-1} & 0 &  0 \\
    0 & 0 & 0 &  0 \\
    0 & 0 & 0 & I_{N-1}
    \end{bmatrix}\mathcal{O}^{-1}.
\end{equation}

Using the formula for $\Theta$ in Eq. \ref{eq:Theta} and the definition of the overlap as a determinant in Eq. \ref{eq:overlaps} we see that
\begin{equation}
    \Theta =\sqrt{\abs{\frac{\det(1+\Gamma[-t\mathcal{D}]\Gamma_{GS})}{\det(1+\Gamma[-t\mathcal{D}]\Gamma_{E})}}}.
\end{equation}

Inserting the expression for $\Gamma[-t\mathcal{D}]$ at long times from the previous line and performing standard determinant manipulations one arrives at the simpler form
\begin{equation}
    \Theta = \frac{\begin{vmatrix}
    v_0 & u_{0j} \\
    v_{i0} & u_{ij} 
    \end{vmatrix}}{\begin{vmatrix}
    v_0 & v_{0j} \\
    u_{i0} & u_{ij} 
    \end{vmatrix}}.
\end{equation}

By introducing the expressions we have for the eigenvectors and rearranging the rows and columns of the determinants we arrive at the expression
\begin{equation}
\Theta = \frac{2\tan mk_0\det\overline{C} \det S}{ \det\overline{S}\det C},
\end{equation}
where
\begin{equation}
    \overline{C} = \begin{bmatrix} 
    1 & 1 & 1 & \dots   & 1 \\
    \cos k_0 & \cos k_1 & \cos k_3 & \dots & \cos k_{N-1} \\
    \vdots & \vdots & \vdots & \ddots & \vdots \\
    \cos mk_0 &  \cos mk_1 & \cos mk_3 & \dots & \cos mk_{N-1}
    \end{bmatrix},
\end{equation}
\begin{equation}
    S = \begin{bmatrix} 
    \sin k_2 & \sin k_4 & \dots   & \sin k_{N-2} \\
    \sin 2k_2 & \sin 2k_4 & \dots & \sin 2k_{N-2} \\
    \vdots & \vdots & \ddots & \vdots \\
    \sin (m-1)k_2 &  \sin (m-1)k_4 & \dots & \sin (m-1)k_{N-2}
    \end{bmatrix},
\end{equation}
\begin{equation}
    C = \begin{bmatrix} 
     1 & 1 & \dots   & 1 \\
    \cos k_1 & \cos k_3 & \dots & \cos k_{N-1} \\
    \vdots & \vdots & \ddots & \vdots \\
    \cos (m-1)k_1 & \cos (m-1)k_3 & \dots & \cos (m-1)k_{N-1}
    \end{bmatrix},
\end{equation}
\begin{equation}
        \overline{S} = \begin{bmatrix} 
    \sin k_0&\sin k_2 & \sin k_4 & \dots   & \sin k_{N-2} \\
    \sin 2k_0 & \sin 2k_2 & \sin 2k_4 & \dots & \sin 2k_{N-2} \\
    \vdots &\vdots & \vdots & \ddots & \vdots \\
    \sin mk_0 &\sin mk_2 &  \sin mk_4 & \dots & \sin mk_{N-2}
    \end{bmatrix}.
\end{equation}

If we add the rows together to form Chebyshev polynomials, the above can be transformed into Vandermonde matrices with determinants given by the well-known results:
\begin{align}
     \det\overline{C} =& 2^{m^2} \prod_{\substack{i<j, \\
0 \text{ or odd}
}} \sin (\frac{k_i-k_j}{2}) \sin (\frac{k_i+k_j}{2}), \\
     \det C =& 2^{(m-1)^2} \prod_{\substack{i<j, \\
\text{odd}
}} \sin (\frac{k_i-k_j}{2}) \sin (\frac{k_i+k_j}{2}), \\
     \begin{split}
     \det S =& 2^{(m-1)^2} \prod_{\substack{i,\\ \text{even}
}} \sin k_i \\
     \times &\prod_{\substack{i<j, \\
\text{even}
}} \sin (\frac{k_i-k_j}{2}) \sin (\frac{k_i+k_j}{2}),
     \end{split}\\
     \begin{split}
     \det \overline{S} =& 2^{m^2} \prod_{\substack{i, \\
0 \text{ or even}
}} \sin k_i \\
     \times&\prod_{\substack{i<j, \\
0 \text{ or even}
}} \sin (\frac{k_i-k_j}{2}) \sin (\frac{k_i+k_j}{2}).
     \end{split}
\end{align}

We can now cancel out the common factors in the numerator and the denominator to obtain
\begin{equation}
    \Theta = 2 \frac{\tan mk_0\prod_{l \text{ odd}} \sin(\frac{k_0-k_l}{2})\sin(\frac{k_0+k_l}{2})}{\sin k_0 \prod_{l \text{ even}} \sin(\frac{k_0-k_l}{2})\sin(\frac{k_0+k_l}{2})},
\end{equation}
which becomes the same as Eq. \ref{eq:Thetaprod} in the main text with the rewriting of the complex wavenumber $K = ik_0$.

\section{Large-$N$ asymptotic expression of entropy}
\label{app:largeN}

In this appendix, we will derive the large-$N$ asymptotic behavior of the exact product expansion of $\Theta$ in Eq. \ref{eq:Thetaprod} of the main text. We start by introducing the auxiliary complex functions
\begin{equation}
\begin{split}
    f(z) &= \frac{g \sin z}{1-g \cos z}, \\
    \phi(z) &= \arctan f(z) = \frac{1}{2i}\log \frac{1-ge^{-iz}}{1-ge^{iz}}.
\end{split}
\end{equation}

In terms of these, we can write the quantization equation as
\begin{equation}
\begin{split}
    &\sin (kN - \phi(k)) = 0, \\
    &k_l = \frac{\pi l}{N} + \frac{\phi(k_l)}{N},
\end{split}
\end{equation}
so that by symmetry $k_0 = 0$ and $k_{-l} = -k_l$. Then we consider the complex function
\begin{equation}
    g(z) = \frac{(N-\phi'(z))\log (\sin(\frac{iK-z}{2}))}{\sin(zN-\phi(z))},
\end{equation}
where
\begin{equation}
    \phi'(z) = \frac{g\cos z-g^2}{1+g^2-2g\cos z}.
\end{equation}

This function is not analytic at the simple poles $z=k_l$ and $z = \pm i\log g$ and at the branch points $z = \pm iK$. Note that $K$ becomes exponentially close to $-\log g$ in the limit of $KN \to \infty$. If we integrate the function around a contour that encircles (but remains close to) the real axis interval $[-\pi,\pi]$ and apply the residue theorem we obtain
\begin{equation}
\begin{split}
    \frac{1}{2\pi i}\oint_C g(z) &dz = \log (\frac{\sin iK}{2}) + \\
    \sum_{l=1}^{N-1} (-1)^l &\log (\sin(\frac{iK-k_l}{2})\sin (\frac{iK+k_l}{2})),
    \end{split}
\end{equation}
where we note that only half the residue at the edge poles is included. Comparing this to the product expansion of $\Theta$ we find that
\begin{equation}
\label{eq:logtheta}
    \log \Theta = \log \tanh \frac{NK}{2} - \frac{1}{2\pi i} \oint_C g(z) dz.
\end{equation}

We are free to deform the integration contour without changing the result of the integral, so long as we do not cross any singularity. The integrals over the sides of the strip are equal to
\begin{equation}
\begin{split}
    \int_{-\infty}^\infty g(\pi+ix) idx + \int_{\infty}^{-\infty} g(-\pi+ix) idx = \\
    2\pi i\int_{-\infty}^{\infty}dx\frac{N+ \frac{g\cosh x+g^2}{1+g^2+2g\cosh x}}{(e^{xN}\sqrt{\frac{1+ge^x}{1+ge^{-x}}}-e^{-xN}\sqrt{\frac{1+ge^{-x}}{1+ge^{x}}})},
    \end{split}
\end{equation}
which cancels because the integrand is an odd function of $x$. We can express the function $g(z)$ explicitly as
\begin{equation}
\begin{split}
    &g(z) = \\
    -2i &e^{izN}\sqrt{\frac{1-ge^{iz}}{1-ge^{-iz}}}\frac{N- \frac{g\cos z-g^2}{(1-ge^{iz})(1-ge^{-iz})}}{1-e^{2izN}\frac{1-ge^{iz}}{1-ge^{-iz}}}\log \sin \frac{iK-z}{2}
\end{split}
\end{equation}

We will focus on the contour integration around the singularity in the top half of the complex plane $z \sim iK$. The integration on the bottom half follows by analogy, but we find that it has no contribution to the order of approximation we are interested in, so we simply discard it. If we make the change of coordinate $z\to iz' = z-i\log \frac{1}{g}$ and consider points sufficiently far from the origin that
\begin{equation}
\label{eq:approxradius}
    \frac{\abs{z'}}{\log \frac{1}{g}} \gg e^{-2N \log \frac{1}{g}},
\end{equation}
along with the additional approximation that $-N\log g \gg 1$, then $g'(z') = g(z)$ is simplified to
\begin{equation}
\begin{split}
    &g'(z') = \\
    -2&e^{-KN - z'N}\sqrt{\frac{1-g^2e^{-z'}}{z'}}\left(N+ \frac{1-g^2}{2(1-g^2e^{-z'}) z'}\right) \\
    &\times \log \sin \frac{z'}{2}.
\end{split}
\end{equation}

In this coordinate, the integration is performed along a dumbbell contour, going along the branch cut on the positive half of the real axis and encircling the singularity at the origin. The radius $\epsilon$ is always understood to obey the condition in Eq. \ref{eq:approxradius}. Since the function decays very fast along the positive real axis for large $N$, we can make a further approximation on our contour
\begin{equation}
    g'(z') =
    -2e^{-KN - z'N}\sqrt{\frac{1-g^2}{z'}}\left(N+ \frac{1}{2 z'}\right)\log \sin \frac{z'}{2},
\end{equation}

where the error we incur is of order $\mathcal{O}(\frac{1}{KN})$ in the integral. We can then express the integral contribution to $\log \Theta$ as
\begin{equation}
\begin{split}
    &\frac{1}{2\pi}\int_C g'(z') dz' = \\
    \frac{1}{\pi}e^{-NK} &\sqrt{1-g^2}\int_{C} \frac{dz}{\sqrt{z}}e^{-Nz}(N+\frac{1}{2z}) \log \frac{z}{2}.
\end{split}
\end{equation}

The circular region has a contribution of
\begin{equation}
\begin{split}
    \int_{l_0} \frac{dz}{\sqrt{z}}e^{-Nz}(N+\frac{1}{2z})\log \frac{z}{2} &= -\frac{2}{\sqrt{\epsilon}} \log \frac{\epsilon}{2}-\frac{4}{\sqrt{\epsilon}} \\
    &= -2\sqrt{\frac{N}{\epsilon'}} (\log \frac{\epsilon'}{2N}+2).
\end{split}
\end{equation}

The 2 linear portions contribute
\begin{equation}
\begin{split}
    \int_{l_++l_-} &\frac{dz}{\sqrt{z}}e^{-Nz}(N+\frac{1}{2z})\log \frac{z}{2} = \\ 
    2\int_\epsilon^\infty &\frac{dx}{\sqrt{x}} e^{-Nx}(N+\frac{1}{2x}) \log \frac{x}{2} = \\
    2\sqrt{N} \int_{\epsilon'}^\infty &\frac{dy}{\sqrt{y}} e^{-y} (1+\frac{1}{2y})\log \frac{y}{2N}.
\end{split}
\end{equation}

The integrals appearing in the formula above have the following values
\begin{align}
    \int_{0}^\infty \frac{dy}{\sqrt{y}}e^{-y} &= \sqrt{\pi}, \\
    \int_{0}^\infty \frac{dy}{\sqrt{y}}e^{-y} \log y &= -\sqrt{\pi}(\gamma+\log 4), \\
    \int_{\epsilon'}^\infty \frac{dy}{y^{\frac{3}{2}}}e^{-y} &= \frac{2}{\sqrt{\epsilon'}} - 2\sqrt{\pi} +\mathcal{O}(\sqrt{\epsilon'}), \\
    \int_{\epsilon'}^\infty \frac{dy}{y^{\frac{3}{2}}}(e^{-y}-1) \log y &= -2\sqrt{\pi}\psi(-\frac{1}{2}) + \mathcal{O}(\sqrt{\epsilon'}), \\
    \int_{\epsilon'}^\infty \frac{dy}{y^{\frac{3}{2}}} \log y &= \frac{2(\log \epsilon'+2)}{\sqrt{\epsilon'}}.
\end{align}

If we put all pieces together we arrive at
\begin{equation}
    \int_{C} \frac{dz}{\sqrt{z}}e^{-Nz}(N+\frac{1}{2z})\log \frac{z}{2} = -4\sqrt{\pi N},
\end{equation}
so we have that
\begin{equation}
    \frac{1}{2\pi i} \oint_{C} g(z) dz = -4\sqrt{\frac{N}{\pi}}e^{-NK} \sqrt{1-g^2}(1 + \mathcal{O}(\frac{1}{NK})).
\end{equation}

If we introduce this into Eq. \ref{eq:logtheta} we get
\begin{equation}
    \log \Theta = 2e^{-NK}(2\sqrt{\frac{N}{\pi}} \sqrt{1-g^2}-1),
\end{equation}
which we introduce in the defining equation of the entropy to get the result provided in the main text.

\bibliography{apssamp}

\end{document}